\documentclass[10pt]{IEEEtran}

\ifCLASSOPTIONcompsoc%
  \usepackage[nocompress]{cite}
\else
  \usepackage{cite}
\fi

\usepackage{amssymb}
\usepackage[english]{babel}
\usepackage[free-standing-units,per-mode=repeated-symbol,mode=text,detect-weight=true,detect-family=true,retain-explicit-plus=true]{siunitx}
\usepackage[nonumberlist]{glossaries}
\usepackage{hyperref}
\usepackage{cleveref}
\usepackage{enumitem}
\usepackage{pgfplots}
\usepackage{pgfplotstable}
\usepackage{tikz}
\usepackage{enumitem}
\usepackage{booktabs}
\usepackage{url}
\usepackage{multirow}
\usepackage[final, activate={true, nocompatibility}, kerning=true, babel=true]{microtype}
\usepackage{xfrac}
\usepackage[inkscapelatex=false]{svg}
\usepackage{xspace}
\usepackage{multicol}

\newacronym[longplural={Scratchpad Memories}]{SPM}{SPM}{Scratchpad Memory}
\newacronym[longplural={Standard Cell Memories}]{SCM}{SCM}{Standard Cell Memory}
\newacronym[longplural={Static Random-Access Memories}]{SRAM}{SRAM}{Static Random-Access Memory}
\newacronym{1RW}{1RW}{single read/write port}
\newacronym{3DIC}{3D-IC}{three-dimensional integrated circuit}
\newacronym{3R1W}{3R1W}{three read ports and one write port}
\newacronym{ACE}{ACE}{AXI Coherent Extensions}
\newacronym{AI}{AI}{Artificial Intelligence}
\newacronym{AMBA}{AMBA}{Advanced Microcontroller Bus Architecture}
\newacronym{APB}{APB}{Advanced Peripheral Bus}
\newacronym{API}{API}{Application Programming Interface}
\newacronym{ASIC}{ASIC}{Application-Specific Integrated Circuit}
\newacronym{AVX}{AVX}{Advanced Vector Extension}
\newacronym{AXI}{AXI}{Advanced eXtensible Interface}
\newacronym{BEOL}{BEOL}{back end of the line}
\newacronym{BLAS}{BLAS}{Basic Linear Algebra Subprograms}
\newacronym{C4}{C4}{controlled collapse chip connection}
\newacronym[longplural={Core Complexes}]{CC}{CC}{Core Complex}
\newacronym{CHI}{CHI}{Coherent Hub Interface}
\newacronym{CMG}{CMG}{Core Memory Group}
\newacronym{CMOS}{CMOS}{Complementary Metal-Oxide-Semi\-con\-ductor}
\newacronym{CNN}{CNN}{Convolutional Neural Network}
\newacronym{CPU}{CPU}{Central Processing Unit}
\newacronym{CSR}{CSR}{Control and State Register}
\newacronym{CTS}{CTS}{Clock Tree Synthesis}
\newacronym{DCT}{DCT}{discrete cosine transform}
\newacronym{DDR}{DDR}{double data rate}
\newacronym{DLP}{DLP}{Data Level Parallelism}
\newacronym{DMA}{DMA}{Direct Memory Access}
\newacronym{DNN}{DNN}{Deep Neural Network}
\newacronym{DRAM}{DRAM}{Dynamic Random-Access Memory}
\newacronym{DRV}{DRV}{design rule violation}
\newacronym{DSP}{DSP}{Digital Signal Processing}
\newacronym{DUT}{DUT}{Device Under Test}
\newacronym{ECL}{ECL}{Emitter-Coupled Logic}
\newacronym{EDP}{EDP}{energy-delay product}
\newacronym{F2F}{F2F}{face-to-face}
\newacronym{FBB}{FBB}{Forward Body-Biasing}
\newacronym{FDSOI}{FD-SOI}{Fully Depleted Silicon on Insulator}
\newacronym{FEOL}{FEOL}{front end of the line}
\newacronym{FFT}{FFT}{Fast Fourier Transform}
\newacronym{FINFET}{FinFET}{Fin Field-Effect Transistor}
\newacronym{FMA}{FMA}{Fused Multiply-Add}
\newacronym{FPGA}{FPGA}{Field-Pro\-gram\-ma\-ble Gate Array}
\newacronym{FPRF}{FPR}{Floating-Point Register File}
\newacronym{FU}{FU}{Functional Unit}
\newacronym{FPU}{FPU}{Floating Point Unit}
\newacronym{GPGPU}{GPGPU}{General-Purpose \acrlong{GPU}}
\newacronym{GPRF}{GPR}{General-Purpose Register File}
\newacronym{GPU}{GPU}{Graphics Processing Unit}
\newacronym{HDL}{HDL}{Hardware Description Language}
\newacronym{HERO}{HERO}{Heterogeneous Embedded Research Platform}
\newacronym{HPC}{HPC}{High-Performance Computing}
\newacronym{IC}{IC}{integrated circuit}
\newacronym{ICG}{ICG}{Integrated Clock Gating}
\newacronym{IDol}{I\$}{Instruction Cache}
\newacronym{ILP}{ILP}{Instruction Level Parallelism}
\newacronym{IOT}{IoT}{Internet of Things}
\newacronym{IPC}{IPC}{Instructions Per Cycle}
\newacronym{IPU}{IPU}{Integer Processing Unit}
\newacronym{ISA}{ISA}{Instruction Set Architecture}
\newacronym{LMUL}{LMUL}{Vector Length Multiplier}
\newacronym{LSU}{LSU}{Load/Store Unit}
\newacronym{LVT}{LVT}{low voltage threshold}
\newacronym{MACU}{MACU}{Multiply-Accumulate Unit}
\newacronym{MIMD}{MIMD}{Multiple Instruction, Multiple Data}
\newacronym{ML}{ML}{Machine Learning}
\newacronym{MMU}{MMU}{Memory Management Unit}
\newacronym{MUL}{MUL}{multiplier}
\newacronym{MVE}{MVE}{M-Profile Vector Extension}
\newacronym{MVL}{MVL}{maximum vector length}
\newacronym{NOC}{NoC}{Network-on-Chip}
\newacronym{NUMA}{NUMA}{non-uniform memory access}
\newacronym{PCIe}{PCIe}{Peripheral Component Interconnect Express}
\newacronym{PC}{PC}{Program Counter}
\newacronym{PDP}{PDP}{power-delay product}
\newacronym{PE}{PE}{Processing Element}
\newacronym{PL}{PL}{Programmable Logic}
\newacronym{PMCA}{PMCA}{Programmable Manycore Accelerator}
\newacronym{PPA}{PPA}{power, performance, and area}
\newacronym{PSL}{PSL}{Power Service Layer}
\newacronym{PTE}{PTE}{page-table entry}
\newacronym{PTW}{PTW}{page-table walker}
\newacronym{PULP}{PULP}{Parallel Ultra Low Power}
\newacronym{RAM}{RAM}{Random-Access Memory}
\newacronym{RAW}{RAW}{read-after-write}
\newacronym{RBB}{RBB}{Reverse Body-Biasing}
\newacronym{ROB}{ROB}{Reorder Buffer}
\newacronym{RTL}{RTL}{Register Transfer Level}
\newacronym{RVT}{RVT}{Regular Voltage Threshold}
\newacronym{RVV}{RVV}{RISC-V Vector Extension}
\newacronym{RoCC}{RoCC}{Rocket Custom Coprocessor Interface}
\newacronym{SDRAM}{SDRAM}{synchronous dynamic random-access memory}
\newacronym{SIMD}{SIMD}{Single Instruction, Multiple Data}
\newacronym{SIMT}{SIMT}{Single Instruction, Multiple Thread}
\newacronym{SLDU}{SLDU}{Slide Unit}
\newacronym{SLVT}{SLVT}{super-low voltage threshold}
\newacronym{SM}{SM}{Streaming Multiprocessor}
\newacronym{SOC}{SoC}{System-on-Chip}
\newacronym{SSE}{SSE}{Streaming SIMD Extension}
\newacronym{SSR}{SSR}{Stream Semantic Register}
\newacronym{STA}{STA}{Static Timing Analysis}
\newacronym{STCO}{STCO}{System-Technology Co-Optimization}
\newacronym{SVE}{SVE}{Scalable Vector Extension}
\newacronym{TLP}{TLP}{Thread Level Parallelism}
\newacronym{TSV}{TSV}{through-silicon via}
\newacronym{TxnID}{TxnID}{Transaction ID}
\newacronym{VAC}{VAC}{Vector Access}
\newacronym{VAU}{VAU}{Vector Arithmetic Unit}
\newacronym{VCONV}{VCONV}{Vector Conversion}
\newacronym{VC}{VC}{virtual channel}
\newacronym{VEX}{VEX}{Vector Execute}
\newacronym{VFU}{VFU}{vector functional unit}
\newacronym{VID}{VID}{Vector Instruction Decode}
\newacronym{VIS}{VISSUE}{Vector Instruction Issue}
\newacronym{VLA}{VLA}{Vector-Length Agnostic}
\newacronym{VLEM}{VLEM}{Vector Lockstep Execution Mode}
\newacronym{VLIW}{VLIW}{Very Long Instruction Word}
\newacronym{VLOOP}{VLOOP}{Vector Loop}
\newacronym{VLR}{VLR}{vector length register}
\newacronym{VLSU}{VLSU}{Vector Load/Store Unit}
\newacronym{VNB}{VNB}{Von Neumann Bottleneck}
\newacronym{VPU}{VPU}{Vector Processing Unit}
\newacronym{VRF}{VRF}{Vector Register File}
\newacronym{VSLDU}{VSLDU}{Vector Slide Unit}
\newacronym{VT}{VT}{vector thread}
\newacronym{W2W}{W2W}{wafer-to-wafer}
\newacronym{WAR}{WAR}{write-after-read}
\newacronym{WAW}{WAW}{write-after-write}

\newcommand{\qtyadj}[2]{\qty[quantity-product={\text{-}}]{#1}{#2}}

\makeatletter
\let\MYcaption\@makecaption
\makeatother
\usepackage[font=footnotesize]{subcaption}
\makeatletter
\let\@makecaption\MYcaption
\makeatother

\definecolor{PULPRed}{HTML}{A8322C}
\definecolor{PULPBlue}{HTML}{1269B0}
\definecolor{PULPGreen}{HTML}{1DB24B}
\definecolor{PULPOrange}{HTML}{F29545}
\definecolor{PULPPurple}{HTML}{910569}
\definecolor{PULPOlive}{HTML}{48592C}
\definecolor{PULPMarine}{HTML}{007996}
\definecolor{PULPGray}{HTML}{ABABAB}
\definecolor{Red}{HTML}{FF0000}

\colorlet{color1}{PULPBlue}
\colorlet{color2}{PULPRed}
\colorlet{color3}{PULPGreen}
\colorlet{color4}{PULPOrange}
\colorlet{color5}{PULPPurple}
\colorlet{color6}{PULPOlive}
\colorlet{color7}{PULPMarine}

\colorlet{colorCore}{PULPRed}
\colorlet{colorMemory}{PULPBlue}
\colorlet{colorInterconnect}{PULPGreen}
\colorlet{colorAccelerator}{PULPOrange}
\colorlet{colorPeripheral}{PULPPurple}
\colorlet{colorAlert}{Red}

\DeclareSIUnit\bits{bits}
\DeclareSIUnit\flop{FLOP}
\DeclareSIUnit\flops{FLOPS}
\DeclareSIUnit\fma{FMA}
\DeclareSIUnit\gate{GE}
\DeclareSIUnit\op{OP}
\DeclareSIUnit\ops{OPS}
\DeclareSIUnit\cycle{cycle}
\DeclareSIQualifier\double{DP}
\DeclareSIQualifier\single{SP}
\DeclareSIQualifier\half{HP}
\DeclareSIQualifier\bytep{BP}
\DeclareSIUnit[quantity-product= ]\percent{\%}

\newcommand\eg{e.g.,\xspace}
\newcommand\ie{i.e.,\xspace}

\newcommand\etal{et\penalty50\ al.\xspace}

\newcommand\spatz[1]{\ensuremath{\text{Spatz}_{{\text{#1}}}}}

\newcommand\change[1]{\textcolor{black}{#1}}

\newcommand\add[1]{\textcolor{black}{#1}}

\newcommand\pchange[1]{\textcolor{black}{#1}}

\newlist{rdescription}{description}{1}
\AtBeginEnvironment{rdescription}{%
\setlist[rdescription]{leftmargin=\dimexpr\eqboxwidth{Des}+\labelsep}}%

\usetikzlibrary{scopes, calc, shapes, arrows, positioning, patterns, pie}
\tikzset{>=latex}

\pgfplotsset{compat=1.18}
\pgfplotsset{width=\linewidth, height=7cm}
\pgfplotsset{every x tick label/.append style={font=\small}}
\pgfplotsset{every y tick label/.append style={font=\small}}

\pgfplotsset{
  /pgf/declare function={
    roof(\x,\b,\p) = (\b * \x < \p) * \b * \x + (\b * \x >= \p) * \p;}}

\usepgfplotslibrary{external}
\usepgfplotslibrary{groupplots}
\usepgfplotslibrary{fillbetween}

\begin{document}

\title{Spatz: Clustering Compact RISC-V-Based Vector Units to Maximize Computing Efficiency}

\author{Matteo~Perotti, Samuel~Riedel, Matheus~Cavalcante, Luca~Benini \thanks{\textcopyright 2025 IEEE.  Personal use of this material is permitted.  Permission from IEEE must be obtained for all other uses, in any current or future media, including reprinting/republishing this material for advertising or promotional purposes, creating new collective works, for resale or redistribution to servers or lists, or reuse of any copyrighted component of this work in other works.}\thanks{}

\thanks{Matteo~Perotti, Samuel~Riedel, and Matheus~Cavalcante are with the Integrated Systems Laboratory (IIS), ETH Zurich, 8092 Zurich, Switzerland. E-mail: \{mperotti, sriedel, matheus\}@iis.ee.ethz.ch.}
\thanks{Luca Benini is with the Integrated Systems Laboratory (IIS), ETH Zurich, 8092 Zurich, Switzerland, and also with the Department of Electrical, Electronic, and Information Engineering (DEI), University of Bologna, 40126 Bologna, Italy. E-mail: lbenini@iis.ee.ethz.ch.}
}

\markboth{}{Spatz}

\maketitle

\bstctlcite{IEEE:BSTcontrol}


\begin{abstract}
The ever-increasing computational and storage requirements of modern applications and the slowdown of technology scaling pose major challenges to designing and implementing efficient computer architectures.
\add{To mitigate the bottlenecks of typical processor-based architectures on both the instruction and data sides of the memory, we present Spatz, a compact \qtyadj{64}{\bit} floating-point-capable vector processor based on RISC-V's Vector Extension Zve64d. Using Spatz as the main Processing Element (PE), we design an open-source dual-core vector processor architecture based on a modular and scalable cluster sharing a Scratchpad Memory (SCM).}
Unlike typical vector processors, whose \glspl{VRF} are hundreds of \unit{\kibi\byte} large, we prove that Spatz can achieve peak energy efficiency with a latch-based \gls{VRF} of only \qty{2}{\kibi\byte}.
An implementation of the Spatz-based cluster in GlobalFoundries' 12LPP process with eight double-precision \glspl{FPU} achieves an \gls{FPU} utilization just \qty{3.4}{\percent} lower than the ideal upper bound on a double-precision, floating-point matrix multiplication.
The cluster reaches \qty{7.7}{\fma\per\cycle}, corresponding to \qty{15.7}{\giga\flops\double} and \qty{95.7}{\giga\flops\double\per\watt} at \qty{1}{\giga\hertz} and nominal operating conditions (TT, \qty{0.80}{\volt}, \qty{25}{\celsius}), with more than \qty{55}{\percent} of the power spent on the \glspl{FPU}.
Furthermore, the optimally-balanced Spatz-based cluster reaches a \qty{95.0}{\percent} \gls{FPU} utilization (\qty{7.6}{\fma\per\cycle}), \qty{15.2}{\giga\flops\double}, and \qty{99.3}{\giga\flops\double\per\watt} (\qty{61}{\percent} of the power spent in the \gls{FPU}) on a 2D workload with a \numproduct{7x7} kernel, resulting in an outstanding area/energy efficiency of \qty{171}{\giga\flops\double\per\watt\per\square\milli\meter}.
At equi-area, \change{the} computing cluster built upon compact vector processors reaches a \qty{30}{\percent} higher energy efficiency than a cluster with the same \gls{FPU} count built upon scalar cores specialized for stream-based floating-point computation.
\end{abstract}


\glsresetall{}


\section{Introduction}
\label{sec:introduction}

\IEEEPARstart{T}{he} pervasiveness of \gls{AI} and \gls{ML} applications triggered an explosion of computational requirements across many application domains. The required computing of the largest \gls{ML} model doubles every \num{3.4}~months, while its parameter count doubles every \num{2.3}~months~\cite{Naffziger2021, Wuu2022}. As a result, large-scale computing systems struggle to keep up with the increasing complexity of such \gls{ML} models. In fact, the performance of the fastest supercomputers only doubles every \num{1.2}~years~\cite{Wuu2022}, while their power budget is capped around \qty{20}{\mega\watt} by infrastructure and operating cost constraints. Furthermore, smart devices running \gls{AI} applications at the \gls{IOT} edge~\cite{GreenWavesGAP92024} are also tightly constrained in their power budget due to battery lifetime \cite{7954016} and passive cooling requirements \cite{8509120}. Therefore, small and large modern computing architectures must \change{aggressively} optimize their compute and data movement energy and delay~\cite{Verma2022}.

\add{To do so, designers often rely on the extreme domain specialization of hardware architectures to boost their energy efficiency and performance at the expense of flexibility. However, the pace at which \gls{AI} models evolve today makes extreme specialization unsustainable, favoring processor-based architectures that easily adapt to novel algorithms, often augmented with hardware that can leverage the \gls{DLP} common to \gls{AI} workloads.}

\add{One of the historical limits of processor-based architectures is the one-word-at-a-time style of programming inherited from the von Neumann computer~\cite{Backus1978}, which contributes to the exceedingly high traffic between memory and processor. 
This is a major issue for present computer architectures since interconnects cannot keep up with transistor scaling~\cite{Brain2016} and \glspl{SRAM} are experiencing a similar drastic scaling slowdown, leading to large and inefficient interconnects and considerable area and energy efficiency overhead when higher memory bandwidth is required.}

\add{Moreover, processor-based architectures have very limited capacity in their \gls{GPRF}, \ie their L0 storage. This limits data reuse in key applications, such as matrix multiplication, whose arithmetic intensity increases when more data elements are stored in the \gls{GPRF}. For example, on RISC-V's RV64I \gls{ISA}, the \gls{GPRF} comprises \num{32} \qtyadj{64}{\bit} registers, \qty{256}{\byte}~\cite{RVBase2019}.}

\add{Vector processor architectures are commonly presented as solutions to the first problem thanks to their \gls{SIMD} execution, in which a single instruction applies the same operation on multiple elements of a vector, decreasing the number of instructions to fetch, decode, and issue and relieving the burden on the instruction memory side. Also, the \gls{VRF} size of vector processors is a design parameter and is decoupled from the datapath width, allowing for an additional independent design knob. Many researchers documented the positive impact of longer vectors (allowed by larger \glspl{VRF}) on the instruction count and energy on the instruction memory side. However, even if longer vectors proved to increase the architecture's memory latency tolerance \cite{10.1145/3624062.3624231}, there is a lack of quantitative analyses on how the \gls{VRF} size impacts the bandwidth on the data memory thanks to the additional data reuse.}

\add{In addition, even though common vector processor architectures try to maximize the energy efficiency of their compute, they commonly rely on scalar cores with complex logic that leverages the applications' \gls{ILP}, practically limiting the overall architecture's efficiency.}

\add{This paper presents Spatz, an open-source \gls{FPU}-ready compact \gls{VPU} based on the \gls{RVV} specification~\cite{RISCV2022}, and use it as the processing element in a multi-\gls{PE} cluster. We couple it with a tiny and lean scalar core~\cite{Zaruba2020}, pursuing higher performance and energy efficiency by leveraging \gls{DLP} and avoiding complex energy-hungry \gls{ILP}-exploitation logic. 
Spatz reaches high energy efficiency through an optimized latch-based \gls{SCM} \gls{VRF} even without resorting to extremely long vectors. With this work, we show how Spatz does not only fight the \gls{VNB} on the instruction side of the memory, but its \gls{VRF} also allows it to decrease the bandwidth requirement on the L1 data \gls{SPM}. 
The contributions of this paper are:}
\begin{itemize}
\item The architecture of Spatz, a parametric, compact \qtyadj{64}{\bit} \gls{VPU} based on the \gls{RVV} version 1.0. Spatz uses a generic accelerator interface, allowing it to work in tandem with any scalar core compatible with this interface (\Cref{sec:spatz-architecture}). \add{To our knowledge, the Spatz cluster is the first open-source multi-core RVV-based vector processor architecture};
\item The design of a physically-driven latch-based \gls{SCM} acting as a \gls{VRF}, and an analysis of its post-implementation energy consumption as a function of its capacity in GlobalFoundries' \qty{12}{\nano\meter} \gls{FINFET} node (\Cref{sec:vector-register-file});
\item An analysis of the energy efficiency of the Spatz-based L1 cluster as a function of its L0 \gls{VRF} capacity. Our analysis also covers typical clusters built with simple scalar \glspl{PE}, and memory streaming solutions (\Cref{sec:shared-l1-cluster});
\end{itemize}
This work significantly extends the initial Spatz publication~\cite{Spatz2022}, where Spatz was a \qtyadj{32}{\bit} \gls{VPU} with no support for floating-point computation. The addition of \qtyadj{64}{\bit} floating-point support fundamentally changes Spatz' target applications and is an argument for lean \glspl{PE} driving large functional units. Furthermore, the discussion concerning the latch-based \gls{SCM} architecture and the design of a shared-L1 cluster based on Spatz are exclusive to this publication. An implementation of the Spatz-based cluster in GlobalFoundries' \qty{12}{\nm} \gls{FINFET} process with eight double-precision \glspl{FPU} achieves an \gls{FPU} utilization just \qty{3.4}{\percent} lower than the ideal upper bound on a double-precision, floating-point \numproduct{64x64} matrix multiplication. The cluster reaches \qty{7.7}{\fma\per\cycle}, corresponding to \qty{15.7}{\giga\flops\double} and \qty{95.7}{\giga\flops\double\per\watt} at \qty{1}{\giga\hertz} and nominal operating conditions (TT, \qty{0.80}{\volt}, \qty{25}{\celsius}), with more than \qty{55}{\percent} of the power spent on the \glspl{FPU}. At the same area, a computing cluster built upon compact vector processors reaches an energy efficiency \qty{30}{\percent} higher than that of a cluster with the same \gls{FPU} count built upon scalar RISC-V-based cores specialized for stream-based floating-point computation. Spatz is open-sourced under a liberal license\footnote{Available at \url{https://github.com/pulp-platform/spatz}}.


\section{Vector Register File}
\label{sec:vector-register-file}

The \gls{VRF} is at the core of any vector processor. It must provide enough bandwidth for the functional units to achieve high utilization. In the RISC-V \gls{ISA}, the vector instruction with the highest bandwidth requirements is the floating-point multiply-accumulate between two vectors, $\mathtt{vfmacc.vv}$ (and its integer equivalent, $\mathtt{vmacc.vv}$)~\cite{RISCV2022}. The \gls{VRF} must handle reading three operands and writing one result per cycle to sustain one \gls{FMA} per cycle. Furthermore, more \gls{VRF} bandwidth is required to execute memory operations concurrently with high-bandwidth multiply-accumulate operations.

Large vector processors are typically coupled with equally complex scalar processors. For example, CVA6, an application-class RV64GC scalar core, consumes \qty{317}{\pico\joule} per operation of a simple dot product kernel, with only \qty{28}{\pico\joule} spent on the actual computation~\cite{Zaruba2019}. Therefore, long vectors are needed to amortize the large energy overhead associated with fetching and dispatching individual instructions with a large application-class (super-)scalar core. Typical large vector units achieve long vectors and high bandwidth through a large multi-banked \gls{VRF}, with five~\cite{Minervini2022} to eight~\cite{Perotti2024} \gls{1RW} \gls{SRAM} banks per \gls{FPU}. However, this approach requires inflexible fine-grained scheduling of the \gls{VRF} bank accesses~\cite{Minervini2022} or an architecture that can handle banking conflicts when multiple operands reside at the same bank~\cite{Perotti2024}. Therefore, \glspl{VRF} are coupled with operand ``queues'' which store the operands until they are all simultaneously available.

Instead of achieving high functional unit utilization through hardware complexity, the RISC-V-based Snitch core tries to tackle the \gls{VNB} by maximizing the cores' compute/control ratio, mitigating the efficiency loss due to deep pipelines and dynamic scheduling~\cite{Zaruba2020}. The Snitch-based shared-L1 cluster couples Snitch cores, typically RV32I or RV32E, with large double-precision \glspl{FPU}. It achieves an energy efficiency of \qty{79}{\giga\flops\double\per\watt} on a double-precision matrix multiplication kernel~\cite{Zaruba2020}, almost double that of state-of-the-art vector machines~\cite{Minervini2022}. This cluster's high performance is linked to \glspl{SSR}~\cite{Zaruba2020}, which stream L1 data into and from the \glspl{FPU} without explicit load/store instructions. As a result, \glspl{SSR} incur high L1 traffic. For example, each core must read two L1 \gls{SPM} words per cycle to sustain the execution of a matrix multiplication kernel without incurring systematic structural hazards. This traffic is particularly taxing on the cluster's physical implementation, as the L1 \gls{SPM} interconnect is the critical factor for its scalability~\cite{Paulin2022}. Furthermore, the breakdown of \gls{SRAM} and interconnect scaling makes achieving the L1 bandwidth required by \glspl{SSR} challenging.

\begin{figure}[htbp]
  \centering \includesvg[width=.85\linewidth]{fig/cluster_vector}
  \caption{A shared-L1 cluster design with $C$ \glspl{PE}, each
    controlling $F$ \glspl{FPU}, and a multi-banked L1
    \change{Scratchpad Memory (\gls{SPM})} with $M$ \gls{SRAM} banks.}
  \label{fig:sharedl1_cluster}
\end{figure}

The vector-\gls{SIMD} \add{\gls{ISA}} provides a clean abstraction for software to adapt automatically to the microarchitecture's vector length~\cite{Nigel2017}. \add{Also, as further discussed in the next sections, the \gls{VRF} increases the data reuse of key applications and relieves the bandwidth requirement on the L1 \gls{SCM} interconnect.} Therefore, we elect a streamlined \gls{VPU} with a small \gls{VRF} to act as the \gls{PE} of \add{a shared-L1 cluster, depicted in} \Cref{fig:sharedl1_cluster}. \add{We choose a shared-L1 cluster as an architectural template because of its modularity and ample adoption from \glspl{SOC} for edge-\gls{AI} \cite{GreenWavesGAP92024} to \Glspl{GPU} \cite{NvidiaH1002020}}.
\add{In our proposed architecture, the} \gls{VRF} size is used as an architectural knob to balance the bandwidth and energy cost of an L1 \gls{SRAM} with the size and energy cost of the L0 \gls{SCM}~\cite{Kung1986}. If the \gls{VPU} achieves a high compute/control ratio, we can explore the capacity/bandwidth trade-off without needing large \glspl{VRF}.

The \gls{VPU} is based on the \gls{RVV} extension version \num{1.0} of the open-source RISC-V \gls{ISA}~\cite{RISCV2022}. The extension adds \num{32} architectural vector registers, $\mathtt{v0}$ to \change{$\mathtt{v31}$}, each with \change{$\mathtt{VLENB}$} bytes. We assume that the \gls{VPU} operates on vector elements \qtyadj{8}{\byte} wide. Therefore, each vector register has $\mathtt{vl} = \sfrac{\change{\mathtt{VLENB}}}{8}$ vector elements. Furthermore, the \gls{RVV} extension adds the concept of a \emph{vector register group} so that a single vector instruction can operate on multiple vector registers. The \gls{LMUL} $\ell$, with $\ell \in \{1, 2, 4, 8\}$, represents how many vector registers are combined to form a vector register group. As a result, \gls{LMUL} trades available vector registers against longer vectors at runtime. Taking it into account, \gls{RVV} provides $\sfrac{32}{\ell}$ vector registers, each with $\sfrac{\ell \times \change{\mathtt{VLENB}}}{8}$ elements.

We implement the \gls{VRF} as a multi-banked, multi-ported, latch-based \gls{SCM}. \add{We consider the exploration of an \gls{SCM}-based memory as strategic due to the slowdown of \gls{SRAM} scaling. In fact, modern \gls{AI} accelerators often use \gls{SCM} \cite{Scherer2022}.} The $32\,\change{\mathtt{VLENB}}$~bytes of the \gls{VRF} are divided into two \gls{SCM} banks with \gls{3R1W} and $16\,\change{\mathtt{VLENB}}$~bytes each. A single \gls{3R1W} \gls{SCM} cut can provide enough bandwidth for executing the $\mathtt{vfmacc.vv}$ instruction and multiple \gls{SCM} banks ensure that other instructions can execute concurrently with the high-bandwidth vector multiply-accumulate instructions. Furthermore, since all operands of a vector instruction can be read simultaneously, we can forgo any ``operand buffers'' to time-multiplex the \gls{VRF} operand fetching.

\begin{figure}[htbp]
  \centering
  \includesvg[width=.85\linewidth]{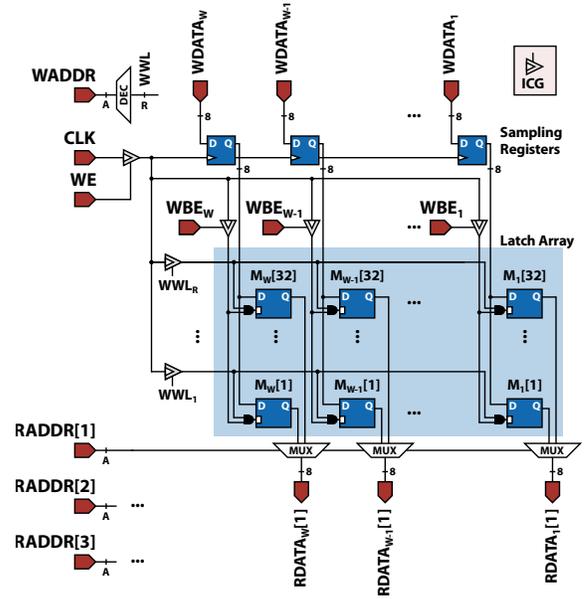}
  \caption{Architecture of a \gls{3R1W} latch-based \gls{SCM} with $R$
    $W$-byte-wide rows, for a total capacity $K = WR$~bytes.}
  \label{fig:vrf_architecture}
\end{figure}

\Cref{fig:vrf_architecture} shows the architecture of a \gls{3R1W} latch-based \gls{SCM} with $R$ rows, each $W$-bytes wide. This \gls{SCM} architecture is similar to that of~\cite{Teman2016, Jun2016}, with manually-inserted \gls{ICG} cells gating the many clocks reaching the latch array. Besides the clock ($\mathtt{CLK}$), write enable ($\mathtt{WE}$), write address ($\mathtt{WADDR}$), and write data ($\mathtt{WDATA}$) ports, the \gls{SCM} also has a write byte strobe ($\mathtt{WBE}$) port to select which bytes are written at each write transaction. The write data is stored in registers before being stored in the latch array. The three read address ($\mathtt{RADDR[i]}$) ports directly control muxes that select a row to forward to the corresponding read data ($\mathtt{RDATA[i]}$) port. Each cell of the storage array comprises eight data latches and an AND2 standard cell, \qty{28}{\gate} in total. For comparison, an equivalent storage array cell based on multi-bit flip-flop standard cells would occupy \qty{49}{\gate}.

\begin{figure}[htbp]
  \centering

  \pgfplotsset{
    /pgf/declare function={
      energy(\w,\k,\a,\b,\c) = (0.001 * \a * \w + 0.001 * \b * \w * \k + 0.001 * \c * \k;}}

  \begin{minipage}[h]{0.5\linewidth}
    \begin{tikzpicture}[every mark/.append style={mark size=3.5pt, ultra thick}]
      \begin{axis}[
        height              = 5.2cm,
        xlabel              = {$W$ [\si{\byte}]},
        xmin                = 8,
        xmax                = 64,
        xtick               = \empty,
        extra x ticks       = {8, 16, 32, 64},
        xmode               = log,
        log basis x         = 2,
        ylabel              = {Read Energy [\si{\pico\joule}]},
        ymin                = 0.1,
        ymax                = 20,
        ymode               = log,
        log basis y         = 10,
        grid                = major,
        legend style        = {at={(0.95,0.05)}, anchor=south east, font=\scriptsize},
        log ticks with fixed point]

        \addlegendimage{only marks, thick, color1, mark=+}
        \addlegendimage{only marks, thick, color2, mark=star}
        \addlegendimage{only marks, thick, color3, mark=x}
        \addlegendentry{$R = \num{64}$}
        \addlegendentry{$R = \num{32}$}
        \addlegendentry{$R = \num{16}$}

        \pgfplotstableread{results/vrf/vrf64}\loadedtable
        \addplot [only marks, thick, mark=+, color1] table [x=Width, y=ReadPower] {\loadedtable};
        \pgfplotstableread{results/vrf/vrf32}\loadedtable
        \addplot [only marks, thick, mark=star, color2] table [x=Width, y=ReadPower] {\loadedtable};
        \pgfplotstableread{results/vrf/vrf16}\loadedtable
        \addplot [only marks, thick, mark=x, color3] table [x=Width, y=ReadPower] {\loadedtable};

        \addplot[color3, dashed, domain=8:64, samples=51]{energy(x, x*16, 47.7588, 0.01792, 0.27497)};
        \addplot[color2, dashed, domain=8:64, samples=51]{energy(x, x*32, 47.7588, 0.01792, 0.27497)};
        \addplot[color1, dashed, domain=8:64, samples=51]{energy(x, x*64, 47.7588, 0.01792, 0.27497)};
      \end{axis}
    \end{tikzpicture}
  \end{minipage}\hfill%
  \begin{minipage}[h]{0.5\linewidth}
    \begin{tikzpicture}[every mark/.append style={mark size=3.5pt, ultra thick}]
      \begin{axis}[
        height              = 5.2cm,
        xlabel              = {$W$ [\si{\byte}]},
        xmin                = 8,
        xmax                = 64,
        xtick               = \empty,
        extra x ticks       = {8, 16, 32, 64},
        xmode               = log,
        log basis x         = 2,
        ylabel              = {Write Energy [\si{\pico\joule}]},
        ymin                = 0.1,
        ymax                = 20,
        ymode               = log,
        log basis y         = 10,
        grid                = major,
        log ticks with fixed point]

        \pgfplotstableread{results/vrf/vrf64}\loadedtable
        \addplot [only marks, thick, mark=+, color1] table [x=Width, y=WritePower] {\loadedtable};
        \pgfplotstableread{results/vrf/vrf32}\loadedtable
        \addplot [only marks, thick, mark=star, color2] table [x=Width, y=WritePower] {\loadedtable};
        \pgfplotstableread{results/vrf/vrf16}\loadedtable
        \addplot [only marks, thick, mark=x, color3] table [x=Width, y=WritePower] {\loadedtable};

        \addplot[color3, dashed, domain=8:64, samples=51]{energy(x, x*16, 72.0772, 0.005721, 3.11102)};
        \addplot[color2, dashed, domain=8:64, samples=51]{energy(x, x*32, 72.0772, 0.005721, 3.11102)};
        \addplot[color1, dashed, domain=8:64, samples=51]{energy(x, x*64, 72.0772, 0.005721, 3.11102)};
      \end{axis}
    \end{tikzpicture}
  \end{minipage}
  \caption{Energy consumption of a \gls{3R1W} latch-based \gls{SCM} with $R$ rows of width $W$ and capacity $K = WR$~bytes. The dashed lines correspond to the functions in \Cref{eq:4,eq:5}.}
  \label{fig:vrf_power}
\end{figure}
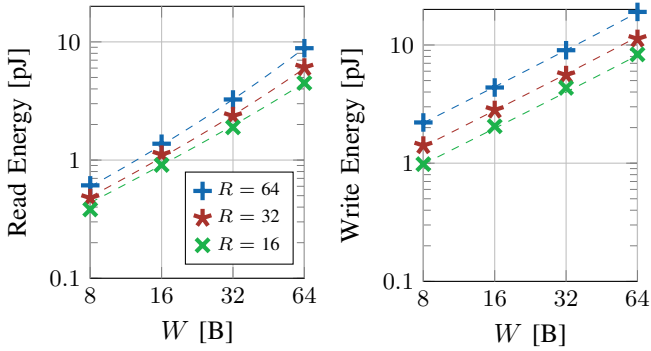

We implemented the \gls{SCM} of \Cref{fig:vrf_architecture} with Synopsys' Fusion Compiler 2022.03, for many combinations of $W$ and $R$, at \qty{950}{\mega\hertz} in worst-case conditions (SS, \qty{0.72}{\volt}, \qty{125}{\celsius}) using GlobalFoundries' 12LPP advanced \qty{12}{\nano\meter} \gls{FINFET} node. Then, we used Synopsys' PrimePower 2022.03 to measure its read and write energy consumption, initialized with random numbers, at \qty{1}{\giga \hertz} and nominal operating conditions (TT, \qty{0.80}{\volt}, \qty{25}{\celsius}). Those results are summarized in \Cref{fig:vrf_power}.

We interpolate the read and write energy consumption of the \gls{SCM} as a polynomial function on the \gls{SCM} width $W$ and capacity $K$. Using the least-squares algorithm to fit parameterized functions to the points of \Cref{fig:vrf_power}, it costs
\begin{equation}
  \label{eq:4}
  \tilde{\varepsilon}^{\text{read}}_{\text{L0}}(W,K) = \num{47.759}W + \num{0.018}WK + \num{0.275}K~\unit{\femto\joule}
\end{equation}
to read $W$~bytes out of the \gls{SCM}, and
\begin{equation}
  \label{eq:5}
  \tilde{\varepsilon}^{\text{write}}_{\text{L0}}(W,K) = \num{72.077}W + \num{0.006}WK + \num{3.111}K~\unit{\femto\joule}
\end{equation}
to write $W$~bytes into the \gls{SCM}. For comparison, it costs $\tilde{\varepsilon}^{\text{read}}_{\text{L1}} = \qty{4.63}{\pico\joule}$ to read and $\tilde{\varepsilon}^{\text{write}}_{\text{L1}} = \qty{5.77}{\pico\joule}$ to write \num{8}~bytes into a \gls{1RW} \gls{SRAM} of capacity \qty{8}{\kibi\byte} in the same technology.

We can gain further insight into the L0 and L1 energy consumption by normalizing the cost per byte accessed. For example, it takes \qty{0.58}{\pico\joule\per\byte} to read data from an \gls{1RW} \gls{SRAM} of width \qty{8}{\byte} and capacity \qty{8}{\kibi\byte}. According to our model, it takes \qty{35}{\percent} less energy, only \qty{0.38}{\pico\joule\per\byte}, to read data from an L0 \gls{SCM} with the same capacity and width. This comparison is not favorable towards the \glspl{SRAM} since our model does not scale to this capacity (the largest \qtyadj{8}{\byte}-wide \gls{SCM} we considered has a capacity of only \qty{512}{\byte}). However, our results back our conclusion that latch-based \glspl{SCM} are an energy-efficient approach for building small-capacity storage buffers. Unfortunately, this \gls{SCM} with a capacity of \qty{8}{\kibi\byte} would be at least $25\times$ larger than its equivalent \gls{SRAM}, which constitutes a major drawback of latch-based \glspl{SCM} with large capacity.

In the following section, we will \add{discuss how to tune the \gls{SCM}-based \gls{VRF} capacity to a target workload} to minimize the overall energy consumption of a shared-L1 computing cluster. In particular, we consider the L1 \gls{SPM} to have a fixed configuration. An extension of this work would be the multi-variate optimization of the cluster's energy consumption as a function of both the L0 \gls{SCM} and L1 \gls{SPM} configurations. This multi-variate analysis requires a model of the \gls{SRAM} energy consumption, which could be inferred from either commercial or open-source memory macro generators.


\section{\add{Matching workload and VRF for optimal efficiency}}
\label{sec:shared-l1-cluster}

\add{Equipped with the efficient \gls{SCM} \gls{VRF} architecture of \Cref{sec:vector-register-file} and its compact energy consumption model, we now shift our focus to the shared-L1 cluster efficiency. 
This section provides a quantitative analysis of the trade-off between the \gls{VRF} (L0) \gls{SCM} capacity and the L1 \gls{SRAM} bandwidth, which influences the architecture's energy efficiency.
Higher \gls{VRF} sizes improve the data reuse of certain applications (reducing their bandwidth requirements) at the expense of higher read/write energy consumption.}

\add{For this study, we generalize the model of Choi~\etal{}\cite{Choi2013}, applying their analysis to our hybrid \gls{SRAM} and \gls{SCM}-based memory hierarchy. 
The presented method can be used to match the \gls{VRF} size with the target workload executed by the architecture. As an example, we discuss the \gls{VRF} sizing considering a target workload dominated by matrix multiplication \footnote{Most of the \gls{DNN} models introduced in the last decade are dominated by matrix multiplication \cite{zhu2024scalablematmulfreelanguagemodeling}.}. We focus our study on the multiplication of two double-precision floating-point matrices, $\mathbf{C} \leftarrow \mathbf{A} \cdot \mathbf{B}$. 
For a more precise tuning on multiple target applications, the analysis can be repeated for every kernel, and the results weighted together to select the optimal \gls{VRF} size.}

\Cref{fig:sharedl1_cluster} shows the architecture of a shared-L1 cluster equipped with compact vector \glspl{PE}. The $C$ \glspl{PE} control $F$ double-precision \glspl{FPU} each. Each \gls{PE} has a \gls{VRF} divided into two \gls{3R1W} \gls{SCM} banks, each $8F$-bytes wide. Since the $F$ double-precision \glspl{FPU} of each \gls{PE} produce $8F$~bytes per cycle, a single \gls{3R1W} \gls{SCM} bank provides enough bandwidth to sustain $F$ \glspl{FMA} per cycle. Furthermore, the L1 \gls{SPM} is implemented as \num{16} \gls{SRAM} banks of \qty{8}{\kibi\byte} each, \qty{128}{\kibi\byte}. All matrices are $n \times n$ and are preallocated in the cluster's \gls{SPM}.

Kung's balance principle states that the machine's inherent balance point should be no greater than the algorithm's arithmetic intensity~\cite{Kung1986}. The architectural balance relationship is generalized by~\cite{Czechowski2011}, considering the machine's degree of memory parallelism and the algorithm's degree of compute parallelism. Applying the relationship to the matrix multiplication kernel on our shared-L1 cluster, we find that
\begin{equation}
  \label{eq:3}
  \frac{CF}{\beta} \le \sqrt{Z},
\end{equation}
where $\beta$ is the \gls{PE}'s bandwidth into the L1 \gls{SPM}, and $Z$ is the total capacity of a \gls{PE}'s \gls{VRF}. Therefore, if we increase the \gls{VRF} capacity by a factor $\alpha$, \ie $Z' = \alpha Z$, we can decrease the L1 bandwidth by a factor $\beta' = \sfrac{\beta}{\sqrt{\alpha}}$ without changing the machine's balance. In other words, we can trade L0 capacity against L1 bandwidth to relax the physical scalability issue of the \gls{PE}-to-L1 interconnect. Furthermore, we can exploit the L0 capacity knob to optimize the cluster's energy efficiency.

\change{Our model assumes that the matrix multiplication is a representative kernel for data-parallel workloads. Ultimately, we will optimize the energy efficiency of the shared-L1 cluster running a matrix multiplication and expect its energy efficiency for other kernels to be near-optimal. Thanks to lean \glspl{PE}, kernels with even higher arithmetic intensity (\eg 2D convolution) will run efficiently thanks to the low overhead of a small controller. Nevertheless, for memory-bound kernels, it still makes sense to have some L0 \gls{SCM}. Namely, when performance and memory bandwidth are well-matched (\ie the kernel is only marginally memory-bound), some L0 storage helps hide the memory latency from upper memory levels. In an alternative direction of exploration for future work, highly memory-bound kernels could bypass the L0 \gls{SCM} completely and operate directly on elements from the L1 \gls{SPM}~\cite{Gobieski2019}.}

\subsection{Energy Consumption Model}
\label{sec:energy-cons-model}

In the following, we will analyze the energy consumption of each major component of the shared-L1 cluster.

\subsubsection{FPUs}
\label{sec:fpus}

Assume that the \glspl{FPU} achieve peak utilization, \ie each executes one \gls{FMA} per cycle. If it costs $\tilde{\varepsilon}_{\text{FPU}}~\unit{\pico\joule}$ for an \gls{FPU} to execute an \gls{FMA} instruction, the $CF$ \glspl{FPU} consume
\begin{equation}
  \label{eq:fpu}
  \varepsilon_{\text{FPU}} = CF\tilde{\varepsilon}_{\text{FPU}}~\unit{\pico\joule\per\cycle}.
\end{equation}

\subsubsection{PEs}
\label{sec:pes}

It costs $\tilde{\varepsilon}'_{\text{PE}}~\unit{\pico\joule}$ for a \gls{PE} to fetch, decode, and dispatch the $i$ instructions of the kernel's hot loop. This loop is \pchange{three} instructions long in a typical vectorized matrix multiplication kernel~\cite{Perotti2024}. Therefore, on average, each \gls{PE} would consume $\tilde{\varepsilon}_{\text{PE}} \triangleq \sfrac{\tilde{\varepsilon}'_{\text{PE}}}{i}~\unit{\pico\joule}$ every cycle to fetch, issue, and dispatch the instructions of the hot loop, as long as the \gls{PE} sustains an \gls{IPC} rate of \num{1}.

The vector-\gls{SIMD} abstraction amortizes this von-Neu\-mann-related bottleneck in two ways. First, the instruction issue cost of each \gls{PE} is amortized by its $F$ \glspl{FPU} running in parallel. Furthermore, since each vector instruction potentially encodes enough micro-operations to keep the \gls{FPU} datapath busy for many cycles, the \gls{PE} can keep the \glspl{FPU} completely utilized at a lower \gls{IPC}. As discussed in \Cref{sec:vector-register-file}, each of the $\sfrac{32}{\ell}$ vector registers has $\sfrac{\ell \times \change{\mathtt{VLENB}}}{8}$ elements. Since the \glspl{FPU} produce $F$ elements per cycle, a vector instruction takes $\sfrac{\ell \times \change{\mathtt{VLENB}}}{8F}$ cycles to execute. In particular, our matrix multiplication implementation uses $\ell = 4$. Therefore, the $C$ \glspl{PE} consume
\begin{equation}
  \label{eq:pe}
  \varepsilon_{\text{PE}} = \tilde{\varepsilon}_{\text{PE}} \frac{2CF}{\change{\mathtt{VLENB}}}~\unit{\pico\joule\per\cycle}.
\end{equation}

\subsubsection{L0 SCM}
\label{sec:l0-scm}

The $F$ \glspl{FPU} of each \gls{PE} must each be able to fetch three double-precision operands out of the \gls{VRF} to achieve a peak collective result throughput of $F$ double-precision results per cycle without any hazards. Since each \gls{SCM} bank is $16\,\change{\mathtt{VLENB}}$~\unit{\byte} large, we can apply \Cref{eq:4,eq:5} to estimate that\pchange{, because of the \gls{FPU} accesses,} the L0 of the $C$ \glspl{PE} consume, on average,
\begin{equation}
  \label{eq:l0}
  \varepsilon_{\text{L0}} = C[3\tilde{\varepsilon}_{\text{L0}}^{\text{read}}(8F, 16\,\change{\mathtt{VLENB}}) + \tilde{\varepsilon}_{\text{L0}}^{\text{write}}(8F, 16\,\change{\mathtt{VLENB}})]~\unit{\pico\joule\per\cycle}.
\end{equation}

\subsubsection{L1 SPM}
\label{sec:l1-spm}

Since the \glspl{FPU} operate on operands stored in the L0 \gls{SCM}, we must account for the data movement cost between the two memory hierarchy levels in the shared-L1 cluster. We will split the analysis into transfers from L0 to L1 and vice-versa. First, regardless of the L0 capacity, we must move $n^2$ elements from the L0 to the L1, corresponding to the $n^2$ elements of matrix $\mathbf{C}$. Assuming that the $n \times n$ matrix multiplication kernel takes $\sfrac{n^3}{CF}$~cycles to execute, \ie we achieve peak \gls{FPU} utilization, it costs
\begin{align}
  \varepsilon_{{\text{L0}} \rightarrow {\text{L1}}} &= \frac{1}{\sfrac{n^3}{CF}} \left[\frac{n^2\tilde{\varepsilon}_{\text{L0}}^{\text{read}}(8F, 16\,\change{\mathtt{VLENB}})}{F} + n^2\tilde{\varepsilon}_{\text{L1}}^{\text{write}}\right]\nonumber\\
                                         &= \frac{C\tilde{\varepsilon}_{\text{L0}}^{\text{read}}(8F, 16\,\change{\mathtt{VLENB}}) + CF \tilde{\varepsilon}_{\text{L1}}^{\text{write}}}{n}~\unit{\pico\joule\per\cycle}  \label{eq:l1_read}
\end{align}
to transfer the matrix multiplication results from the L0 to the L1.
Furthermore, we can use \Cref{eq:3} to estimate the cost of L1 to L0 transfers as a function of the L0 \gls{SCM} capacity. We assume that an \gls{FPU} needs at least eight $8$-\unit{\byte} wide registers to achieve full utilization, a total capacity of $64$~bytes. This accounts for four registers to store intermediate accumulations (since our \gls{FMA} pipeline has four cycles of latency), and four registers to hold matrix operands. In this case, each \gls{FPU} requires $2$ words per cycle out of the L1 \gls{SPM}, at a cost $\varepsilon_{\text{L1}}^{\text{read}} = 2 \tilde{\varepsilon}_{\text{L1}}^{\text{read}}~\unit{\pico\joule\per\cycle}$. Therefore, considering that each \gls{VRF} has a capacity of $32\,\change{\mathtt{VLENB}}$~bytes, the $C$ \glspl{PE} spend
\begin{equation}
  \label{eq:l1_write}
  \varepsilon_{{\text{L1}} \rightarrow {\text{L0}}} = C \left[ \frac{2F \tilde{\varepsilon}_{\text{L1}}^{\text{read}} + 2\tilde{\varepsilon}_{\text{L0}}^{\text{write}}(8F, 16\,\change{\mathtt{VLENB}})}{\sqrt{\sfrac{32\,\change{\mathtt{VLENB}}}{64}}} \right]~\unit{\pico\joule\per\cycle}
\end{equation}
copying elements from the L1 memory into the L0 \gls{VRF}. Therefore,
data transfers between the L0 and the L1 memories consume
$\varepsilon_{\text{L1}} = \varepsilon_{{\text{L0}} \rightarrow
  {\text{L1}}} + \varepsilon_{{\text{L1}} \rightarrow
  {\text{L0}}}~\unit{\pico\joule\per\cycle}$.

\subsection{Energy Efficiency Optimization}
\label{sec:energy-effic-optim}

Assume a shared-L1 cluster implementation with $C = 2$ \glspl{PE}, each controlling $F = 4$ \glspl{FPU}. The eight \glspl{FPU} per cluster are similar to equivalent Snitch-based clusters~\cite{Paulin2022}. \pchange{Based on an explorative implementation of the Snitch scalar core in 12-nm technology}, we estimate that it costs \pchange{$\tilde{\varepsilon}_{\text{PE}} = \qty{3.6}{\pico\joule}$} for Snitch to fetch, decode, and dispatch an instruction of the matrix multiplication kernel and $\tilde{\varepsilon}_{\text{FPU}} = \qty{13.3}{\pico\joule}$ for an \gls{FPU} to execute a double-precision \gls{FMA} instruction. We can use \Cref{eq:fpu,eq:pe,eq:l0,eq:l1_read,eq:l1_write} to estimate the energy consumption of the shared-L1 cluster as a function of the vector length $\change{\mathtt{VLENB}}$, as seen in \Cref{fig:energy_consumption}.

\pgfmathsetmacro\eFPU{13.31}
\pgfmathsetmacro\ePE{3.6}
\pgfmathsetmacro\eSPMread{4.63}
\pgfmathsetmacro\eSPMwrite{5.77}

\pgfplotsset{
  /pgf/declare function={
    eL0read(\w,\k) = 0.001 * (47.7588*\w + 0.01792*\w*\k + 0.27497*\k);}}
\pgfplotsset{
  /pgf/declare function={
    eL0write(\w,\k) = 0.001 * (72.0772*\w + 0.005721*\w*\k + 3.11102*\k);}}

\pgfplotsset{
  /pgf/declare function={
    eFPU(\c,\f,\v,\g) = \c * \f * \eFPU * (1 + \g);}}
\pgfplotsset{
  /pgf/declare function={
    ePE(\c,\f,\v,\g) = (2 * \c * \f * \ePE / \v) * (1 + \g);}}
\pgfplotsset{
  /pgf/declare function={
    eL0(\c,\f,\v,\g) = (3 * \c * eL0read(8*\f, 16*\v) + \c * eL0write(8*\f, 16*\v)) * (1 + \g);}}
\pgfplotsset{
  /pgf/declare function={
    eL0L1(\c,\f,\v) = (\c * eL0read(8*\f, 16*\v) + \c * \f * \eSPMwrite) / 256;}}
\pgfplotsset{
  /pgf/declare function={
    eL1L0(\c,\f,\v) = (2 * \c * \f * \eSPMread + 2 * \c * eL0write(8*\f, 16*\v)) * sqrt(2 / \v);}}
\pgfplotsset{
  /pgf/declare function={
    eL1(\c,\f,\v,\g) = (eL0L1(\c, \f, \v) + eL1L0(\c, \f, \v)) * (1 + \g);}}

\pgfplotsset{
  /pgf/declare function={
    eCluster(\c,\f,\v,\ga,\gb,\gc,\gd) = eFPU(\c, \f, \v, \ga) + ePE(\c, \f, \v, \gb) + eL0(\c, \f, \v, \gc) + eL1(\c, \f, \v, \gd);}}

\begin{figure}[htbp]
  \centering
  \begin{tikzpicture}[every mark/.append style={mark size=3.5pt, ultra thick}]
    \begin{axis}[
      height              = 5.5cm,
      width               = 7.0cm,
      xlabel              = {$\change{\mathtt{VLENB}}$ [\si{\byte}]},
      xmin                = 8,
      xmax                = 256,
      xmode               = log,
      log basis x         = 2,
      ylabel              = {Energy [\si{\pico\joule\per\cycle}]},
      ymin                = 0,
      ymax                = 200,
      grid                = major,
      legend pos          = outer north east,
      log ticks with fixed point]

      \addlegendimage{thick, colorMemory}
      \addlegendimage{thick, colorInterconnect}
      \addlegendimage{thick, colorCore}
      \addlegendimage{thick, colorAccelerator}
      \addlegendentry{$\varepsilon_{\text{L1}}$}
      \addlegendentry{$\varepsilon_{\text{L0}}$}
      \addlegendentry{$\varepsilon_{\text{PE}}$}
      \addlegendentry{$\varepsilon_{\text{FPU}}$}

      \addplot[colorAccelerator, thick, domain=8:256, samples=51, name path=rFPU]{eFPU(2, 4, x, 0)};
      \addplot[colorCore, thick, domain=8:256, samples=51, name path=rPE]{eFPU(2, 4, x, 0) + ePE(2, 4, x, 0)};
      \addplot[colorInterconnect, thick, domain=8:256, samples=51, name path=rVRF]{eFPU(2, 4, x, 0) + ePE(2, 4, x, 0) + eL0(2, 4, x, 0)};
      \addplot[colorMemory, thick, domain=8:256, samples=51, name path=rTCDM]{eCluster(2, 4, x, 0, 0, 0, 0)};

      \path[name path=xAxis] (axis cs:8,0) -- (axis cs:256,0);
      \addplot[fill opacity = 0.25, colorAccelerator] fill between [of=xAxis and rFPU];
      \addplot[fill opacity = 0.25, colorCore] fill between [of=rFPU and rPE];
      \addplot[fill opacity = 0.25, colorInterconnect] fill between [of=rPE and rVRF];
      \addplot[fill opacity = 0.25, colorMemory] fill between [of=rVRF and rTCDM];
    \end{axis}
  \end{tikzpicture}
  \caption{Breakdown of the energy consumption per cycle of the
    shared-L1 cluster, as a function of its vector length
    $\change{\mathtt{VLENB}}$.}
  \label{fig:energy_consumption}
\end{figure}
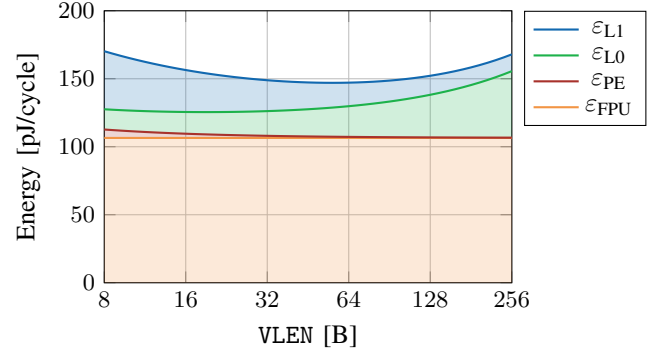

The energy breakdown highlights the benefit of a lightweight \gls{PE}. The \glspl{FPU} dominate the cluster's energy consumption, altogether responsible for about \qty{60}{\percent} of its energy consumption. On the other hand, Snitch is responsible for less than \qty{1}{\percent} of the cluster's energy consumption per cycle. Therefore, longer vectors do little to amortize the cluster's \gls{VNB} since the instruction dispatch overhead is negligible. Furthermore, the breakdown also highlights the data movement overhead and the balance between L0 and L1 energy consumption, which together amount to \qty{30}{\percent} of the cluster's energy consumption.

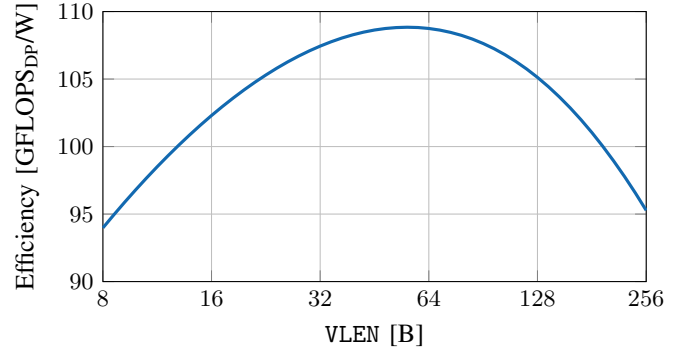
\begin{figure}[htbp]
  \centering
  \resizebox{\linewidth}{!}{\begin{tikzpicture}[every mark/.append style={mark size=3.5pt, ultra thick}]
    \begin{axis}[
      height              = 5.2cm,
      xlabel              = {$\change{\mathtt{VLENB}}$ [\si{\byte}]},
      xmin                = 8,
      xmax                = 256,
      xmode               = log,
      log basis x         = 2,
      ylabel              = {Efficiency [\si{\giga\flops\double\per\watt}]},
      ymin                = 90,
      ymax                = 110,
      grid                = major,
      legend style        = {at={(0.95,0.05)}, anchor=south east, font=\scriptsize},
      log ticks with fixed point]

      \addplot[color1, very thick, domain=8:256, samples=51, name path=rTCDM]{16 * 1000 / eCluster(2, 4, x, 0, 0, 0, 0)};
    \end{axis}
  \end{tikzpicture}}
  \caption{Energy efficiency $\Phi$ of the cluster executing a
    $256 \times 256$ matrix multiplication kernel, as a function of
    the vector length $\change{\mathtt{VLENB}}$.}
  \label{fig:efficiency}
\end{figure}

We can exploit the L0/L1 energy consumption balance to optimize the cluster's energy efficiency $\Phi$ while running the matrix multiplication kernel. As seen in \Cref{fig:efficiency}, the cluster reaches a peak energy efficiency of \pchange{\qty{106.9}{\giga\flops\double\per\watt}} for a vector length of \qty{47}{\byte}. \pchange{At the closest power-of-two with the best energy efficiency}, the cluster reaches \pchange{\qty{106.4}{\giga\flops\double\per\watt}} for a vector length of \qty{64}{\byte}\pchange{, with just a \qty{0.04}{\textperthousand} deviation from the maximum}.

Assuming $\change{\mathtt{VLENB}} = \qty{64}{\byte}$, each \gls{VRF} is \qty{2}{\kibi\byte} large, divided into two \gls{SCM} banks with \num{32} rows of width \qty{32}{\byte}, \qty{1}{\kibi\byte} large. Therefore, the $C = 2$ \glspl{VRF} account for \qty{4}{\kibi\byte} of L0 storage, compared to the \qty{128}{\kibi\byte} of L1 \gls{SPM} storage divided into \num{16} \gls{SRAM} banks. In this scenario, the \glspl{FPU} consume $\varepsilon_{\text{FPU}} = \qty{106.5}{\pico\joule\per\cycle}$, \pchange{the \glspl{PE} consume $\varepsilon_{\text{PE}} = \qty{0.9}{\pico\joule\per\cycle}$,} the \glspl{VRF} consume \pchange{$\varepsilon_{\text{L0}} = \qty{25.7}{\pico\joule\per\cycle}$} \pchange{on \glspl{FPU} accesses, the data transfers between the \glspl{VRF} and the L1 \gls{SPM} \gls{SRAM} banks consume} $\varepsilon_{\text{L1}} = \qty{17.3}{\pico\joule\per\cycle}$, \pchange{and the total energy consumed by \glspl{VRF} and L1 memory banks every cycle is \qty{29.8}{\pico\joule\per\cycle} and \qty{13.3}{\pico\joule\per\cycle}, respectively}. Furthermore, each vector instruction takes \num{8} cycles to execute.

Applying \Cref{eq:3}, we expect the \glspl{PE} to fetch about \num{3} L1 words per cycle. Since the cluster has $M = 16$ \gls{SRAM} banks, we have enough bandwidth for a \gls{DMA} engine to concurrently load data from an external L2 memory into the L1 \gls{SPM} while the \glspl{PE} execute their computations.

\change{We can assess our model's robustness through a sensitivity analysis. Namely, we assess how much the $\mathtt{VLENB}$ that minimizes the cluster's energy efficiency fluctuates in response to \pchange{relative} changes to the model parameters. \pchange{In this analysis, we used 10\% changes, and reported the} most important sensitivity parameters in \Cref{tab:sensitivity}. For example, since $\varepsilon_{\text{FPU}}$ does not depend on $\mathtt{VLENB}$, the vector length that maximizes the cluster's energy efficiency does not depend on $\tilde{\varepsilon}_{\text{FPU}}$.}

\begin{table}[h]
  \centering
  \caption{Sensitivity of the energetically-optimal shared-L1 cluster's $\mathtt{VLENB}$ to the model's parameters. $\alpha$, $\beta$, and $\gamma$ are the coefficients of \Cref{eq:4} and \Cref{eq:5}.}
  \begin{tabular}[h]{ll|ll}
    \toprule
    Parameter                                        & \pchange{Sensitivity$_{10\%}$}    & Parameter & \pchange{Sensitivity$_{10\%}$}                 \\\midrule
    $\tilde{\varepsilon}_{\text{FPU}}$                 & \qty{+0.00}{\byte} & \pchange{$\tilde{\beta}^{\text{read}}_{\text{L0}}$} & \pchange{\qty{-0.80}{\byte}} \\
    
    $\tilde{\varepsilon}_{\text{PE}}$                  & \pchange{\qty{+0.39}{\byte}} & \pchange{$\tilde{\gamma}^{\text{read}}_{\text{L0}}$}  & \pchange{\qty{-0.40}{\byte}} \\
    
    $\tilde{\varepsilon}_{{\text{L1}}}^{\text{read}}$  & \pchange{\qty{+2.40}{\byte}} & \pchange{$\tilde{\alpha}^{\text{write}}_{\text{L0}}$} & \pchange{\qty{+0.30}{\byte}} \\[0.25ex]
    $\tilde{\varepsilon}_{{\text{L1}}}^{\text{write}}$ & \qty{+0.00}{\byte} & \pchange{$\tilde{\beta}^{\text{write}}_{\text{L0}}$}  &  \pchange{\qty{-0.11}{\byte}} \\
    \pchange{$\tilde{\alpha}^{\text{read}}_{\text{L0}}$}          & \pchange{\qty{+0.00}{\byte}} & \pchange{$\tilde{\gamma}^{\text{write}}_{\text{L0}}$}  & \pchange{\qty{-1.71}{\byte}} \\[0.25ex]
    \bottomrule
  \end{tabular}
  \label{tab:sensitivity}
\end{table}

\add{The quantitative conclusions of this section are based on an idealized vector-based shared-L1 cluster whose workload is dominated by matrix multiplications. The proposed \gls{VRF} sizing method can be applied to different applications to show the energy efficiency trade-off caused by varying the arithmetic intensity of the application and the read/write energy of the \gls{VRF}. Based on this analysis, applications with no data reuse or whose data reuse does not depend on their tiling will, therefore, prefer lower \gls{VRF} sizes. However, a \gls{VRF} can also be useful to better tolerate memory latency, especially in architectures with a higher-latency memory hierarchy (this is not the case for our L1-based cluster).}

\add{Throughout the remainder of this publication, we will build out a shared-L1 cluster, starting with the design of the lean Spatz \gls{VPU} as its \gls{PE}. We size Spatz's \gls{VRF} as just discussed due to the centrality of the matrix multiplication kernel and to provide a practical verification of the numbers hypothesized in this Section for our theoretical analysis.}

\section{Spatz: A Compact Vector Processing Unit}
\label{sec:spatz-architecture}

Spatz is a compact parametric \gls{VPU} based on the \acrfull{RVV} version 1.0. \Cref{fig:arch_spatz} shows the microarchitecture of \spatz{F}, a Spatz instance with $F$ \glspl{FPU}, and its integration with a shared-L1 cluster. Snitch and Spatz form a \gls{CC} integrated within a shared-L1 cluster with $16$ \gls{SRAM} banks of \qty{8}{\kibi\byte} each. This section describes Spatz' architecture, highlighting its main components.

\begin{figure}[htbp]
  \centering
  \includesvg[width=\linewidth]{fig/spatz}
  \caption{Microarchitecture of the Spatz-based \gls{CC}, containing a Spatz instance with $F$ \glspl{FPU}, $G$ \glspl{IPU}, and a Snitch scalar core. Snitch and Spatz communicate through CORE-V's X-Interface (XIF). Each Spatz instance has $F+1$ \qtyadj{64}{\bit} memory interfaces.}
  \label{fig:arch_spatz}
\end{figure}

\subsection{Instruction Dispatch}
\label{sec:instruction-dispatch}

Spatz implements the \gls{RVV} \gls{ISA}, version 1.0~\cite{RISCV2022}.
We target the $\mathtt{Zve64d}$ subset, designed for embedded vector machines with 8, 16, 32, and \qtyadj{64}{\bit} integer and floating-point support. Furthermore, we also provide a Spatz configuration targeting the $\mathtt{Zve32x}$ \gls{RVV} subset, with support for 8, 16, and \qtyadj{32}{\bit} integer operations, presented in~\cite{Spatz2022}. Spatz is processor-agnostic and communicates with the scalar core through CORE-V's X-Interface accelerator interface~\cite{OpenHW2022}. Therefore, Spatz can be used with any core compatible with the X-Interface. Snitch's small footprint is particularly adapted for an execution paradigm where most of the computation happens in the vector unit.

Since the accelerator interface specification is still in its infancy, it needs to be better adapted to the memory bandwidth requirements of a vector machine. We extended the X-Interface to consider cases where the accelerator makes its memory access through a wider memory interface than the scalar cores' one. Furthermore, we guarantee the ordering between Spatz' and Snitch's memory requests by stalling the scalar core's \gls{LSU} while Spatz' \gls{VLSU} executes a memory operation and vice-versa.

\subsection{Controller}
\label{sec:controller}

Snitch only pre-decodes vector instructions, dispatching the vector instruction and any scalar operands to the vector unit. Spatz' controller decodes the vector instructions, keeps track of their execution, and acknowledges their completion with Snitch.
The controller also manages the \glspl{CSR} of the \gls{RVV} \gls{ISA}. For example, the $\mathtt{vlen}$ \gls{CSR} defines the vector length of all vector instructions. Another important \gls{CSR} is $\mathtt{vtype}$, which controls the vector elements' width and the vector register grouping \gls{LMUL}. Finally, the controller orchestrates the execution of the vector instructions in the functional units. The scoreboard keeps track of the element-wise progression of each vector instruction. Hazards between vector instructions are handled through operand backpressure. In addition, Spatz supports vector chaining on an element basis.

Within the controller, the \gls{FPU} Sequencer manages the scalar \gls{FPRF}. It implements a simple scoreboard to manage access to instructions that access the \gls{FPRF}, \eg vector instruction with scalar floating-point operands and scalar floating-point instructions. Furthermore, the sequencer also manages scalar floating-point memory operations, hence its dedicated \qtyadj{64}{\bit} memory interface. The \gls{FPU} sequencer is only instantiated when Spatz is configured with \gls{FPU} support.

\subsection{Vector Register File}
\label{sec:vector-register-file-1}

The \gls{VRF} is the core of any vector machine. We implemented Spatz' \gls{VRF} as a multi-banked multi-ported register file with two \gls{3R1W} banks. Each bank is implemented as a latch-based \gls{SCM}. The \gls{VRF} is centralized and serves all functional units. Its ports match the throughput requirements of the $\mathtt{vfmacc}$ instruction, which reads three double-precision operands to produce one double-precision result. Each bank is $\change{\mathtt{VLENB}}/2$ bytes wide, and each of the \num{32} $\change{\mathtt{VLENB}}$-byte vector registers occupies one row in each of the two \gls{VRF} banks. Each \gls{VRF} port is $64F$-\unit{\bit} wide. We studied Spatz' \gls{VRF} in \Cref{sec:vector-register-file}.

The centralized \gls{VRF} helps the implementation of vector instructions with irregular access patterns, \eg vector slides ($\mathtt{vd}[i] \leftarrow \mathtt{vs}[i \pm \mathtt{shamt}]$) and reductions ($\mathtt{vd}[0] \leftarrow \Sigma_i \mathtt{vs}[i]$). Namely, a lane-based vector architecture, where the vector registers of the \gls{VRF} are divided into lanes based on their index $i$, would need extra logic to shuffle the elements and store them in the correct lane, with important scalability implications~\cite{Perotti2024}.

\subsection{Functional Units}
\label{sec:functional-units}

Spatz has three functional units: the \gls{VLSU}, the \gls{VAU}, and the \gls{VSLDU}.

\subsubsection{Vector Arithmetic Unit}
\label{sec:vect-arithm-unit}

The \gls{VAU} is Spatz' main functional unit, hosting $F$ transprecision \glspl{FPU}~\cite{Mach2021} and $G$ \glspl{IPU}. Each \gls{FPU} supports fp8, fp16, fp32, and fp64 computation. Each \gls{IPU} supports \qtyadj{8}{\bit}, \qtyadj{16}{\bit}, \qtyadj{32}{\bit}, and \qtyadj{64}{\bit} computation. All functional units maintain a throughput of \qty{64}{\bit\per\cycle}, regardless of the element width. Spatz decouples the number of integer $G$ and floating-point $F$ functional units. Therefore, we can tune Spatz to focus on integer or floating-point workloads. This paper will analyze the case where $G = 1$. Therefore, Spatz mainly focuses on floating-point workloads, with the integer datapath mainly used for address computations. Furthermore, we also reuse Spatz' datapath to execute some scalar instructions by reinterpreting scalar multiplications, integer divisions, and floating-point operations as vector instructions of unit length that commit into the \gls{GPRF}. \change{Our workloads did not have enough of those scalar instructions to deserve a dedicated functional unit, as derived from our performance results. Thanks to this datapath reuse, we can reduce the scalar core area and energy overhead. A leaner scalar core has a smaller $\tilde\varepsilon_{\text{PE}}$, reducing the vector length that reaches maximal energy efficiency.}

Our \glspl{FPU} include the ExSdotp extension of~\cite{Bertaccini2022}. This \gls{FPU} computational unit implements the widening \emph{sum-of-dot-products} operation. It takes four operands expressed in $w$-bits and an accumulator input $2w$-bits wide to compute a sum of dot products also $2w$-bits wide,
\begin{equation}
  \label{eq:sdotp}
  \text{ExSdotp}_{2w} = {\text{a}}_w {\text{b}}_w + {\text{c}}_w {\text{d}}_w + {\text{e}}_{2w}.
\end{equation}
This operation is particularly useful for \gls{ML} applications. In this scenario, we achieve a smaller memory footprint thanks to the narrow operand width while retaining high precision thanks to the wide accumulation result. \change{Our widening and sum-of-dot-products floating-point engines} \pchange{address} \change{the accuracy loss associated with accumulating low-bitwidth numbers encoded with low-precision formats. Therefore, our workloads target $w = \qty{8}{\bit}$ and $w = \qty{16}{\bit}$, although we have the hardware to accumulate \qtyadj{32}{\bit} elements into a \qtyadj{64}{\bit} accumulator.}

Our previous publication~\cite{Spatz2022} analyzed Spatz' performance as an integer \gls{VPU}. However, since integer computations are energetically cheaper than floating-point computations, data movement is responsible for a larger fraction of the cluster's overall energy consumption. Therefore, the analysis of this paper is a ``worst-case scenario'' concerning the benefits of vector processing since the cluster's power consumption is dominated by the \glspl{FPU}, as seen in \Cref{fig:energy_consumption}.

\subsubsection{Vector Load/Store Unit}
\label{sec:vlsu}

The \gls{VLSU} handles the memory interfaces of Spatz, with support for unit-strided, constant-strided, and indexed memory accesses. The \gls{VLSU} supports a parametric number of \qtyadj{64}{\bit} wide memory interfaces. By default, the number of memory interfaces $F$ matches the number of \glspl{FU} in the design. This implies a peak operation per memory bandwidth ratio of \SI{0.25}{\flop\double\per\byte}.

Spatz' independent and narrow memory interfaces allow the reuse of the same \qtyadj{64}{\bit} wide L1 \gls{SPM} interconnect used by the scalar cores. The independent interfaces also allow fast execution of constant-strided and scatter-gather memory operations, as the \gls{VLSU} does not need to coalesce requests into wide memory transfers. However, since the L1 \gls{SPM} interconnect does not guarantee the ordering between the responses of the individual memory requests, a \gls{ROB} sits between the memory interfaces and the \gls{VRF}. The \gls{ROB} ensures that the memory responses are written in order to the \gls{VRF}, simplifying Spatz' vector chaining mechanism.

\subsubsection{Vector Slide Unit}
\label{sec:vector-slide-unit}

The \gls{VSLDU} executes vector permutation instructions. Examples of such instructions include vector slide up/down and vector moves. The unit operates on two private $64F$-bit-wide register banks. Between those two register banks, an all-to-all interconnect allows the implementation of any permutation scheme. Common operations, \eg slides, produce results at $64F$ bits per cycle ratio, which matches Spatz' other functional units' peak throughput. The register banks also play a role similar to the \gls{ROB} of Spatz' \gls{VLSU}. The \gls{VSLDU} double-buffers on those registers and ensures that the unit commits to the \gls{VRF} in $64F$-bit-wide words, simplifying the chaining calculation in the scoreboard.


\section{Performance Analysis}
\label{sec:performance-analysis}

\begin{figure}[htbp]
  \centering
  \includesvg[width=.85\linewidth]{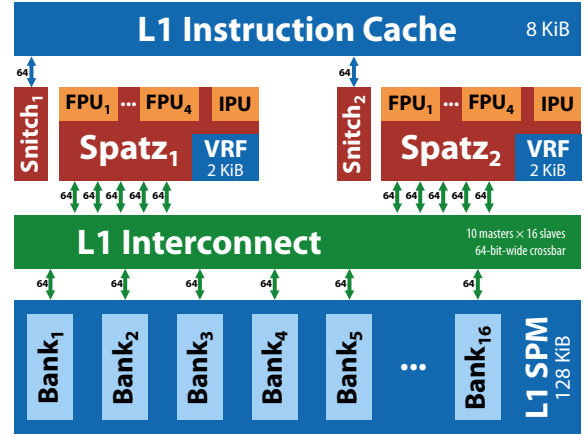}
  \caption{A shared-L1 cluster with two Spatz \glspl{PE}, each
    controlling four multi-precision \glspl{FPU} and one \gls{IPU},
    and a multi-banked L1 \gls{SPM} with \qty{128}{\kibi\byte}.}
  \label{fig:sharedl1_cluster_spatz}
\end{figure}

We implemented a shared-L1 cluster with two Spatz-based \glspl{CC}, each controlling four \glspl{FPU} and an \gls{IPU}. The cluster has eight double-precision \glspl{FPU}, \qty{4}{\kibi\byte} of L0 \gls{VRF}, \qty{8}{\kibi\byte} of L1 \gls{IDol}, and \qty{128}{\kibi\byte} of \gls{SPM} divided into \num{16} banks. The cluster's architecture can be seen in \Cref{fig:sharedl1_cluster_spatz}.

The Spatz-based cluster's performance metrics were then evaluated using a set of common data-parallel workloads,
\begin{description}
\item[matmul:] Matrix multiplication between two $n \times n$ double-precision floating point matrices;
\item[wid-matmul$_w$:] Widening matrix multiplication of two $n \times n$ matrices of element width $w$ into a matrix of width $2w$;
\item[conv2d:] 2D convolution of a double-precision floating point matrix of size $n \times n$ with a kernel of size $7 \times 7$.
\item[dotp:] Scalar product between two double-precision, floating-point vectors of length $n$;
\item[fft:] Implementation of Cooley-Tukey's \gls{FFT} algorithm on a vector with $n$ complex double-precision floating point samples.
\end{description}
All kernels operate on data residing in the cluster's L1 memory. Furthermore, all workloads operate on double-precision floating-point elements, except wid-matmul$_w$, which targets low-precision floating-point formats instead. The performance results were extracted with a cycle-accurate \gls{RTL} simulation of the target workloads. \Cref{tab:performance_results} summarizes the cluster's performance and \gls{FPU} utilization.

\begin{table}[htbp]
  \centering
  \caption{Spatz cluster's multi-core performance and \gls{FPU}
    utilization.}
  \begin{tabular}[h]{rrll}
    \toprule
    Benchmark                                   & $n$  & Perf. [\unit{\flop\per\cycle}] & Util. [\unit{\percent}] \\\midrule
    \multirow{3}{*}{\textbf{matmul}}            & 16   & \num{11.57}                    & \num{72.3}              \\
                                                & 32   & \num{15.00}                    & \num{93.8}              \\
                                                & 64   & \num{15.67}                    & \num{97.9}              \\\midrule
    \multirow{2}{*}{\textbf{wid-matmul}$_{16}$}  & 64   & \num{57.53}                    & \num{89.9}              \\
                                                & 128  & \num{61.52}                    & \num{96.1}              \\\midrule
    \multirow{2}{*}{\textbf{wid-matmul}$_{8}$}   & 64   & \num{112.9}                    & \num{88.2}              \\
                                                & 128  & \num{121.8}                    & \num{95.2}              \\\midrule
    \multirow{2}{*}{\textbf{conv2d}}            & 32   & \num{14.91}                    & \num{93.2}              \\
                                                & 64   & \num{15.20}                    & \num{95.0}              \\\midrule
    \multirow{2}{*}{\textbf{dotp}}              & 256  & \num{1.67}                     & \num{10.4}              \\
                                                & 4096 & \num{5.45}                     & \num{34.0}              \\\midrule
    \multirow{2}{*}{\textbf{fft}}               & 128  & \pchange{\num{3.43}}                     & \pchange{\num{34.2}}              \\
                                                & 256  & \pchange{\num{4.01}}                     & \pchange{\num{40.1}}              \\\bottomrule
  \end{tabular}
  \label{tab:performance_results}
\end{table}

We will assess the \gls{PPA} of the Spatz-based cluster against a baseline scalar cluster using eight Snitch cores as \glspl{PE}. Snitch is a single-issue core, and its instruction issue rate limits the performance of the scalar shared-L1 cluster. Therefore, we will compare all cluster metrics against a specialized Snitch-based cluster implementing \glspl{SSR}, a stream-based \gls{ISA} extension. In particular, RISC-V floating point instructions access the \gls{FPRF}, while most bookkeeping instructions access the \gls{GPRF} instead~\cite{RVBase2019}. \glspl{SSR} exploit this to implement an energy efficient pseudo-double-issue execution between floating-point and integer instructions~\cite{Zaruba2020}. With hardware loops to remove branches and the elision of explicit memory load and store instructions, the Snitch-based \gls{SSR} cluster is highly competitive with the Spatz cluster. \add{The \gls{SSR} Snitch cluster is, therefore, much more efficient than a vanilla scalar Snitch cluster, as \glspl{SSR} with hardware loops can stream vectors of data to/from the cores by means of a few dedicated instructions.} In a high-level sense, the main difference between those clusters is the number of ports in the L1 interconnect. \add{From an application perspective, the \gls{SSR} cluster executes eight scalar RISC-V threads boosted with \gls{SSR} instructions, while the Spatz cluster deals with two \gls{RVV} threads.} Due to its small \gls{GPRF} capacity, each \gls{SSR}-capable Snitch core has three memory ports (\num{24} initiators in total) into the \gls{SPM} interconnect, all connected to \num{32} \gls{SRAM} banks. In contrast, Spatz' \gls{VRF} allows for more data reuse and, consequently, for a reduction in L1 \gls{SPM} bandwidth. Each Spatz \gls{CC} has four ports into the L1 \gls{SPM} (eight ports in total), all connected to \num{16} \gls{SRAM} banks. \change{We also note that \gls{RVV} is a standard extension with compiler support and widespread community usage, unlike \glspl{SSR}, which are a non-standard \gls{ISA} extension.}

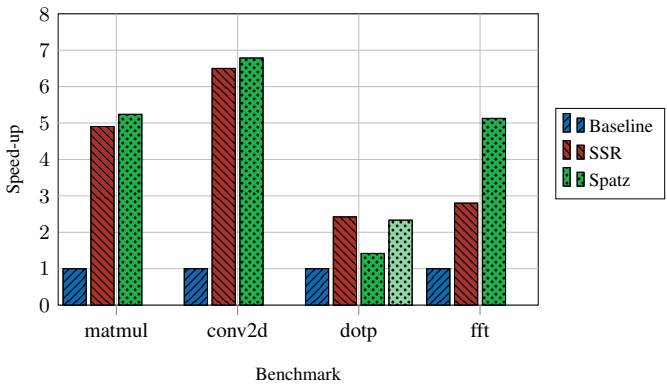
\begin{figure}[htbp]
  \centering
  \resizebox{\linewidth}{!}{\begin{tikzpicture}[/tikz/font=\footnotesize]
    \begin{axis}[
      ybar,
      height = 6cm,
      enlargelimits = 0,
      scaled y ticks = false,
      legend style={
        at={(1.15,0.52)},
        anchor=center,
        legend columns=1
      },
      legend cell align={left},
      ylabel = {Speed-up},
      ymajorgrids = true,
      yminorgrids = true,
      minor y tick num = 0,
      ytick style = {draw=none},
      ymin = 0,
      ymax = 8,
      extra y ticks={1, 3, 5, 7},
      xlabel = {Benchmark},
      xmajorgrids = true,
      xtick pos = left,
      xtick = \empty,
      extra x ticks={0, 1, 2, 3},
      extra x tick labels={matmul, conv2d, dotp, fft},
      enlarge x limits = 0.16]

      \pgfplotstableread{results/spatz/snitch}\loadedtable
      \addplot[fill=color1, postaction={pattern=north east lines}] table [x=Index, y expr=\thisrowno{3}/\thisrowno{4}] {\loadedtable};

      \pgfplotstableread{results/spatz/ssr}\loadedtable
      \addplot[fill=color2, postaction={pattern=north west lines}] table [x=Index, y expr=\thisrowno{3}/\thisrowno{4}] {\loadedtable};

      \pgfplotstableread{results/spatz/spatz}\loadedtable
      \addplot[fill=color3, postaction={pattern=crosshatch dots}] table [x=Index, y expr=\thisrowno{3}/\thisrowno{4}] {\loadedtable};

      \pgfplotstableread{results/spatz/spatzWide}\loadedtable
      \addplot[fill=color3!50, postaction={pattern=crosshatch dots}] table [x=Index, y expr=\thisrowno{3}/\thisrowno{4}] {\loadedtable};

      \legend{Baseline, \gls{SSR}, Spatz}
    \end{axis}
  \end{tikzpicture}}
  \caption{Bar plot of Spatz and \gls{SSR}-based cluster's speed-ups versus the Snitch baseline cluster on a set of data-parallel workloads. All clusters have the same number of \glspl{FPU} and, therefore, the same peak achievable performance. The fourth, lighter bar in the \emph{dotp} kernel indicates the speedup of a Spatz cluster with twice as many \gls{VLSU} interfaces in the L1 \gls{SPM}.}
  \label{fig:perf_comparison}
\end{figure}

\Cref{fig:perf_comparison} shows the speed-up of the Spatz-based and \gls{SSR}-based clusters versus the Snitch-based baseline cluster on a set of data-parallel workloads. For example, Spatz achieves \qty{15.46}{\flop\double\per\cycle} (\qty{96.6}{\percent} of the theoretical peak performance) running a \numproduct{64x64} matrix multiplication. This is $\num{5.2}\times$ higher than the baseline performance, \qty{2.99}{\flop\double\per\cycle}. Meanwhile, the \gls{SSR}-based cluster achieves \qty{14.67}{\flop\double\per\cycle}, a $\num{4.9}\times$ speed-up versus the baseline. A similar trend is also seen for the \emph{conv2d} kernel, with the Spatz-based and \gls{SSR}-based achieving $\num{6.8}\times$ and $\num{6.5}\times$ speed-ups versus the baseline Snitch-based cluster, respectively.

We also achieve high performance with the \emph{wid-matmul}$_{16}$, a widening matrix multiplication that operates on \qtyadj{16}{\bit} floating point operands and accumulates in a \qtyadj{32}{\bit} register. In this scenario, the \qtyadj{64}{\bit} ExSdotp~\cite{Bertaccini2022} datapath of each \gls{FPU} can execute four half-precision \glspl{FMA} per cycle. For a large \numproduct{128x128} matrix multiplication, Spatz reaches \qty{61.5}{\flops\half\per\cycle}, an \gls{FMA} utilization of \qty{96.1}{\percent}. In comparison, the \gls{SSR}-based cluster achieves \qty{52.0}{\flops\half\per\cycle} on the same kernel, an \gls{FPU} utilization \qty{15.5}{\percent} lower than that of the Spatz cluster. Similar trends are seen for lower precision, \eg \emph{wid-matmul}$_8$.

In general, \gls{SSR}'s performance drop on the \emph{matmul} and \emph{conv2d} kernels is explained due to banking conflicts in the L1 \gls{SPM}, which degrade the performance of \gls{SSR}-based solutions due to their high L1 \gls{SPM} bandwidth requirements. \change{Namely, since \glspl{SSR} have very little data reuse, each \gls{SSR}-powered Snitch core reads two elements and writes one element back to the L1 \gls{SPM} per cycle. This implies that \num{24} out of the \num{32} L1 \gls{SPM} \gls{SRAM} banks are used per cycle}. However, the \gls{SSR}-based cluster performs better on the \emph{dotp} kernel. The Spatz-based cluster reaches a \pchange{$\num{1.44}\times$} speed-up versus the baseline, while the \gls{SSR}-based cluster reaches a speed-up \pchange{close to $\num{3}\times$} versus the baseline. This is due to \emph{dotp}'s very low data reuse\change{, which forces to} fetch two \qtyadj{64}{\bit} elements from the L1 \gls{SPM} for each \gls{FMA} operation. The \gls{SSR}-based cluster's large L1 \gls{SPM} interconnect \change{and three memory ports per core} can supply this throughput \change{to the \glspl{FPU}}. On the other hand, the smaller L1 interconnect of the Spatz cluster can provide a single \qtyadj{64}{\bit} element per cycle to each \gls{FPU}, throttling the execution of the \emph{dotp} kernel.

\change{It is important to mention that the performance difference between Spatz and the SSR cluster on low data-reuse kernels, \eg \emph{dotp}., primarily depends on the current Spatz configuration and micro-architecture. 
A Spatz cluster with $2F$ L1 \gls{SPM} interfaces per \gls{VLSU} and streamlined reduction logic can achieve a \emph{dotp} speedup similar to the \gls{SSR}-based cluster while still featuring a lower number of ports in the L1 \gls{SPM} interconnect (\num{16} versus \num{24}), as shown in \Cref{fig:perf_comparison} by the lighter bar relative to the \emph{dotp} kernel. On the other hand, we should note that low-compute intensive kernels like \emph{dotp} are often bottlenecked at the} \pchange{cluster memory interface or at the off-chip memory controller,} \change{weakening the motivation for the local complexity increase.}

The \emph{fft} kernel \change{exemplifies} the cluster's performance on workloads \change{with non-linear access patterns in the L1 Memory}. The \gls{SSR}-based cluster achieves up to \pchange{\qty{1.91}{\flops\double\per\cycle}} on an \gls{FFT} with \num{128} double-precision complex samples, \pchange{$\num{3.2}\times$ faster} than the baseline cluster. The low speed-up is due to the frequent synchronization between the \gls{FFT} stages~\cite{Zaruba2020}. On the other hand, the Spatz-based cluster achieves \pchange{\qty{3.43}{\flops\double\per\cycle}} on the same problem, \pchange{$5.8\times$} faster than the baseline. Spatz benefits from a fast scatter-gather execution mechanism (\Cref{sec:vlsu}). Furthermore, since many \gls{FFT} butterflies execute within a Spatz core, there is much less need for intra-Spatz atomic synchronization.


\section{Implementation Results}
\label{sec:impl-results}

This section analyzes the \gls{PPA} figures of merit of a Spatz-based shared-L1 cluster with key data-parallel workloads.

\subsection{Methodology}
\label{sec:methodology-1}

We used Synopsys Fusion Compiler 2022.03 to synthesize, place, and route the cluster with GlobalFoundries' 12LPP \qty{12}{\nano\meter} advanced \gls{FINFET} node. We target a minimum operating frequency of \qty{950}{\mega\hertz} under worst-case conditions (SS, \qty{0.72}{\volt}, \qty{125}{\celsius}). Furthermore, we used Synopsys PrimeTime 2022.02 for sign-off \gls{STA} and power estimation under nominal conditions (TT, \qty{0.80}{\volt}, \qty{25}{\celsius}) at \qty{1}{\giga\hertz} using switching activities extracted from gate-level simulations.

\subsection{Area Analysis and Breakdown}
\label{sec:area-analysis}

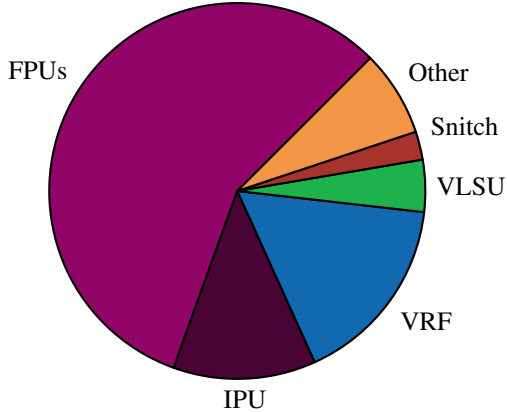
\begin{figure}[ht]
  \centering
  \begin{tikzpicture}[scale=0.8]
    \pie [color = {colorPeripheral, colorPeripheral!50!black, colorMemory, colorInterconnect, colorCore, colorAccelerator}, sum=auto, hide number, radius=2.2, text=label, rotate=45] {
      568/\glspl{FPU},
      126/\gls{IPU},
      169/\gls{VRF},
      46/\gls{VLSU},
      25/Snitch,
      76/Other
    }
  \end{tikzpicture}
  \caption{Post-implementation area distribution of the Spatz-based
    \gls{CC}, \qtyadj{1.01}{\mega\gate}-large. The slices correspond
    to (A) \glspl{FPU}, \change{\qty{568}{\kilo\gate}}; (B) \gls{IPU},
    \qty{126}{\kilo\gate}; (C) \gls{VRF}, \qty{169}{\kilo\gate}; (D)
    \gls{VLSU}, \qty{46}{\kilo\gate}; (E) Snitch,
    \qty{25}{\kilo\gate}; (O) other smaller blocks, \eg \gls{VSLDU},
    \gls{FPU} sequencer, and controller, \qty{76}{\kilo\gate}.}
  \label{fig:spatz_area}
\end{figure}

Each Spatz-based \gls{CC} is about \qty{1.01}{\mega\gate} large. \Cref{fig:spatz_area} shows a post-implementation area breakdown of this \gls{CC}, highlighting its main blocks. The \gls{VAU} is the largest block, with its \qty{694}{\kilo\gate} occupying \qty{69}{\percent} of the \gls{CC}'s area. Within the \gls{VAU}, each \gls{FPU} is \qty{142}{\kilo\gate} large, with the remaining \qty{126}{\kilo\gate} being occupied by the \gls{IPU}---particularly by its \qtyadj{64}{\bit} multiplier---and by the vector reduction logic. The \gls{VRF} occupies \qty{169}{\kilo\gate}, \qty{17}{\percent} of the \gls{CC}'s area. The remaining blocks have small contributions to the \gls{CC}'s area, with Snitch's \qty{25}{\kilo\gate}, in particular, occupying only \qty{2.5}{\percent} of the \gls{CC}'s overall footprint.

\begin{figure}[htbp]
  \centering
  \includesvg[width=.85\linewidth]{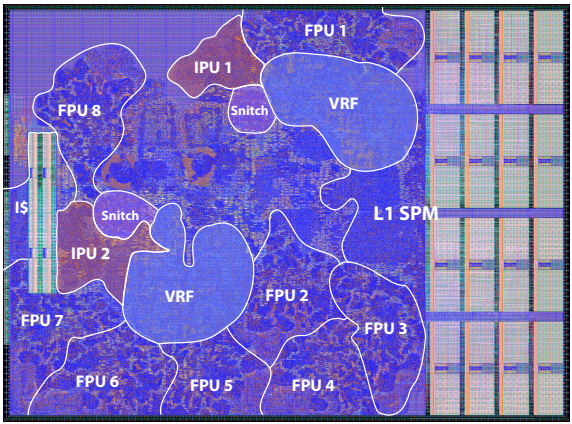}
  \caption{Placed-and-routed Spatz-based shared-L1 cluster,
    implemented as a \qtyproduct{737x1003}{\micro\meter} block. The
    cluster's main blocks are highlighted: namely the Snitch cores,
    \glspl{VRF}, \glspl{IPU}, \glspl{FPU}, L1 \gls{SPM}, and
    \gls{IDol}.}
  \label{fig:layout}
\end{figure}

The cluster was implemented as a block of dimensions \qtyproduct{737x1003}{\micro\meter}, \change{\qty{0.74}{\milli\meter\squared}} in total. \Cref{fig:layout} shows the placed-and-routed Spatz-based shared-L1 cluster, highlighting its main hierarchical blocks. The cluster was implemented with an overall standard cell density of \qty{52}{\percent}. The L1 \gls{SPM} interconnect and the \gls{VRF} are routing-intensive blocks and achieve a lower utilization of the standard cell area. For example, while highly-computational blocks such as the \glspl{FPU} achieve a standard cell utilization of \qty{60}{\percent}, the \glspl{VRF} are placed and routed at a standard cell utilization of \qty{49}{\percent}.

The critical path of the Spatz cluster starts at Snitch's L0 \gls{IDol}, through the Snitch core and Spatz' instruction decoder. This critical path is about $45$-gates long. Furthermore, another long path starting at the L0 \gls{IDol} and through Snitch (without leaving the core) is about the same length. Therefore, Spatz' inclusion does not limit the cluster operating frequency, which is the same as that of the scalar Snitch-based cluster~\cite{Paulin2022}.

\subsection{Energy Breakdown}
\label{sec:energy-breakdown}

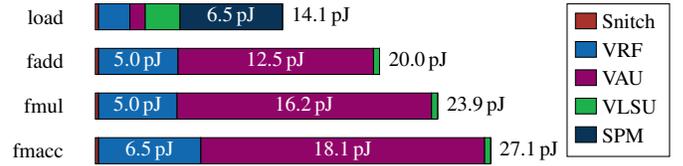
\begin{figure}[ht]
  \centering
  \resizebox{\linewidth}{!}{\begin{tikzpicture}[/tikz/font=\footnotesize]
    \begin{axis}[
      xbar stacked,
      xmax=31.5,
      bar width = 1em,
      y axis line style = {draw=none},
      axis x line = none,
      tickwidth = 0pt,
      height = 3.75cm,
      nodes near coords,
      legend style={at={(1.15,1)}, anchor=north east, cells={anchor=west}},
      symbolic y coords = {fmacc, fmul, fadd, load},
      ytick distance = 1,
      yticklabel style={align=right, xshift=2ex, font=\footnotesize},
      every pin/.style={font=\footnotesize},
      nodes near coords align={center},
      point meta = explicit symbolic]

      \addplot+[xbar, color=black, fill=colorCore] plot coordinates {(0.19,fmacc) (0.19,fmul) (0.19,fadd) (0.19,load)};
      \addplot+[xbar, color=black, text=white, fill=colorMemory] plot coordinates {(6.53,fmacc) [\qty{6.5}{\pico\joule}] (5.01,fmul) [\qty{5.0}{\pico\joule}] (5.07,fadd) [\qty{5.0}{\pico\joule}] (2.01,load)};
      \addplot+[xbar, color=black, text=white, fill=colorPeripheral] plot coordinates {(18.10,fmacc) [\qty{18.1}{\pico\joule}] (16.24,fmul) [\qty{16.2}{\pico\joule}] (12.48,fadd) [\qty{12.5}{\pico\joule}] (0.97,load)};
      \addplot+[xbar, color=black, fill=colorInterconnect] plot coordinates {(0.39,fmacc) (0.39,fmul) (0.39,fadd) (2.22,load)};
      \addplot+[xbar, color=black, text=white, fill=colorMemory!50!black] plot coordinates {(0,fmacc) (0,fmul) (0,fadd) (6.54,load) [\qty{6.5}{\pico\joule}]};

      \addplot+[color=black, nodes near coords align={right}] plot coordinates {(0.0001,fmacc) [\qty{27.1}{\pico\joule}] (0.0001,fmul) [\qty{23.9}{\pico\joule}] (0.0001,fadd) [\qty{20.0}{\pico\joule}] (0.0001,load) [\qty{14.1}{\pico\joule}]};

      \legend{Snitch, \gls{VRF}, \gls{VAU}, \gls{VLSU}, \gls{SPM}}
    \end{axis}
  \end{tikzpicture}}
  \caption{Breakdown of the Spatz-based cluster's energy consumption
    per elementary operation of several vector instructions.}
  \label{fig:energy}
\end{figure}

We measured the energy consumption per elementary operation of the Spatz-based cluster in nominal operating conditions at \qty{1}{\giga\hertz}, using switching activities extracted from a gate-level simulation. \Cref{fig:energy} shows Spatz' energy breakdown when running the $\mathtt{vload}$ (load), $\mathtt{vfadd}$ (floating-point addition), $\mathtt{vfmul}$ (floating-point multiplication), and $\mathtt{vfmacc}$ (floating-point multiply-accumulate) double-precision instructions.

Snitch consumes only \qty{0.19}{\pico\joule} per elementary operation while issuing instructions to the vector unit. The low energy requirement is because Snitch only needs to issue an instruction every \num{32} cycles to keep Spatz' \glspl{FPU} fully utilized. For the computational operations, the \glspl{FPU} are responsible for most of Spatz' energy consumption, up to \qty{67}{\percent} of the cluster's overall energy consumption when running the $\mathtt{vfmacc}$ instruction, followed by the \gls{VRF}, which is responsible for \qty{24}{\percent}. Also of note is the reduction in \gls{VRF} energy consumption between the $\mathtt{vfadd}$ and $\mathtt{vfmul}$ instructions, which read two operands and write one result to the \gls{VRF} per cycle, and the $\mathtt{vfmacc}$ instruction, which reads an additional third operand.

Each \gls{FPU} of the Spatz cluster consumes \qty{18.1}{\pico\joule} to execute an \gls{FMA} operation while executing the $\mathtt{vfmacc}$ instruction. This $\varepsilon_{\text{FPU}}$ value is \qty{38}{\percent} higher than what we estimated from~\cite{Zaruba2020}'s results, lowering our cluster's expected energy efficiency. Adding packed-\gls{SIMD} support and including low-precision floating-point formats justify and explain the efficiency drop. The \gls{FPU} implementation of~\cite{Mach2021}, which has packed-\gls{SIMD} vector support, consumes \qty{26.1}{\pico\joule} to execute a double-precision \gls{FMA} operation, in line with what we measure in Spatz.

\change{It is important to note that the numbers of \Cref{fig:energy} are not directly comparable with the previous Spatz publication~\cite{Spatz2022} due to the widening of its datapath from \qty{32}{\bit} to the current \qty{64}{\bit}. Furthermore, Spatz did not include \glspl{FPU}, which are the} \pchange{focus of this work.}

\subsection{Power Consumption and Energy Efficiency}
\label{sec:power-cons-energy}

\begin{figure}[htbp]
  \centering
  \begin{tikzpicture}[scale=0.7]
    \pie [color = {colorPeripheral, colorMemory, colorInterconnect, colorMemory!50!black, colorInterconnect!50!black, colorCore!40!black, colorCore, colorAccelerator}, sum=auto, hide number, radius=2.5, text=label, rotate=45] {
      87.3/\glspl{FPU},
      31.7/\gls{VRF},
      7.47/\gls{VLSU},
      4.25/\gls{SPM},
      10.69/\gls{SPM} Interconnect,
      9.32/Controller,
      5.65/Snitch,
      9.15/Other
    }
  \end{tikzpicture}
  \caption{Average power consumption breakdown of the Spatz-based cluster running a \numproduct{64x64} double-precision floating-point matrix multiplication. At \qty{1}{\giga\hertz}, the cluster achieves \qty{15.7}{\giga\flops\double} and consumes, on average, \qty{164}{\milli\watt} to execute the workload at nominal operating conditions (TT, \qty{0.80}{\volt}, \qty{25}{\celsius}). The slices correspond to (A) \glspl{FPU}, \qty{87}{\milli\watt}; (B) \gls{VRF}, \qty{34}{\milli\watt}; (C) \gls{VLSU}, \qty{7.5}{\milli\watt}; (D) L1 \gls{SPM} memories, \qty{4.25}{\milli\watt}; (E) L1 \gls{SPM} interconnect, \qty{10.69}{\milli\watt}; (F) Spatz' controller, \qty{10.3}{\milli\watt}; (G) Snitch, \qty{5.6}{\milli\watt}; (O) other smaller hierarchical blocks, \qty{9.1}{\milli\watt}.}
  \label{fig:spatz_power}
\end{figure}
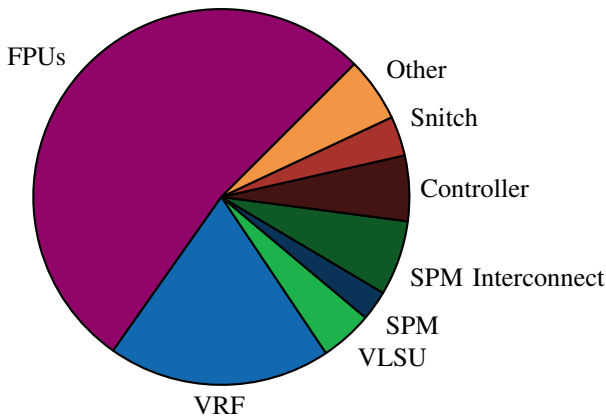

The Spatz cluster consumes \qty{164}{\milli\watt} to execute the \numproduct{64x64} \emph{matmul} kernel. Therefore, the Spatz cluster achieves \qty{15.7}{\giga\flops\double} running the \numproduct{64x64} \emph{matmul}, for an energy efficiency of \qty{95.7}{\giga\flops\double\per\watt}. \Cref{fig:spatz_power} shows a breakdown of the power consumption of the cluster's hierarchical blocks.

The eight \glspl{FPU} consume \qty{87}{\milli\watt}, \qty{54}{\percent} of the cluster's overall power consumption. In comparison, the \glspl{FPU} consumed only \qty{41}{\percent} of the power consumption of the equivalent Snitch-based cluster~\cite{Zaruba2020}. Furthermore, the \gls{VRF} consumes \qty{32}{\milli\watt} (\qty{19}{\percent}). This power consumption ratio mirrors the analysis from \Cref{sec:energy-effic-optim}. However, we underestimated the \gls{VRF} energy consumption by \qty{30}{\percent}, probably due to timing pressure in the \gls{VRF} interface with the \glspl{FU}. Spatz' \gls{VLSU} consumes \qty{7.0}{\milli\watt} (\qty{4.2}{\percent}), and the L1 \gls{SPM} \glspl{SRAM} and interconnect consume \qty{15.4}{\milli\watt} (\qty{8.0}{\percent}). Finally, Spatz' controller consumes \qty{9.3}{\milli\watt} (\qty{5.8}{\percent}) and the Snitch cores and \gls{IDol} consume \qty{5.6}{\milli\watt} (\qty{3.4}{\percent}).

\subsection{Model Validation}
\label{sec:model-validation}

\pchange{As shown in \Cref{tab:validation}, our post-implementation energy measurements correlate with the numbers used in \Cref{sec:shared-l1-cluster}. For example, the \gls{VRF} consumes \qty{34}{\pico\joule\per\cycle}, \qty{4.2}{\pico\joule\per\cycle} (\qty{14}{\percent}) more than the model estimate. We notice a large relative error in the \gls{PE} energy consumption value that corresponds to an absolute error of only \qty{0.8}{\pico\joule\per\cycle}. Furthermore, the relative discrepancies between the hypothesized and measured \gls{FPU}, L0 \gls{VRF}, and L1 \gls{SPM} energy consumption are all less than \qty{20}{\percent}, testifying to the soundness of the shared-L1 cluster's energy consumption model.}

\begin{table}[h]
  \centering
  \caption{Discrepancies between the hypothesized energy consumption in \Cref{sec:shared-l1-cluster} and the post-implementation measured energy consumption.}
  \begin{tabular}[h]{rcccc}
    \toprule
          & Hypothesis [\unit{\pico\joule}] & Measured [\unit{\pico\joule}] & Abs. Error [\unit{\pico\joule}] & Rel. Error \\\midrule
    FPU   & \pchange{\num{106.5}}                & \num{87}           & \pchange{-19.5}               & \pchange{-18\%}                            \\
    PE    & \pchange{\num{0.9}}                  & \pchange{\num{1.7}}                        & \pchange{+0.8} & \pchange{+89\%}                            \\
    L0    & 29.8                 & \num{34.0}  & \pchange{+4.2}                     & \pchange{+14\%}                            \\
    L1    & \num{13.3}                 & \num{15.0} & \pchange{+1.7}                      & \pchange{+13\%}                     
    \\\bottomrule
  \end{tabular}
  \label{tab:validation}
\end{table}

\subsection{PPA comparison with the SSR-based cluster}
\label{sec:comp-with-state}

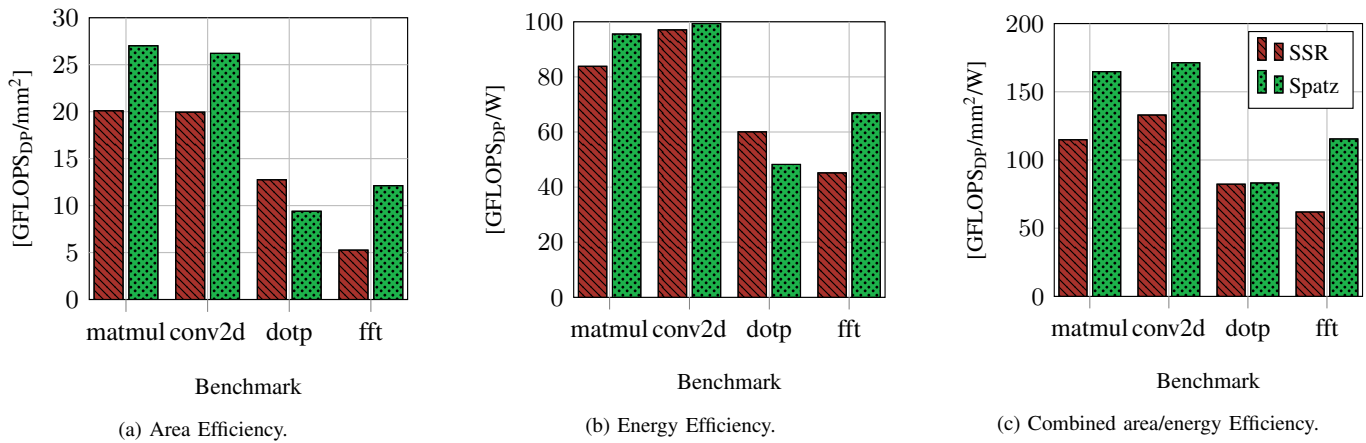
\begin{figure*}
  \centering
  \begin{minipage}[h]{0.3\linewidth}
    \resizebox{\linewidth}{!}{\begin{tikzpicture}[/tikz/font=\footnotesize]
        \begin{axis}[
          ybar,
          height = 5cm,
          enlargelimits = 0,
          scaled y ticks = false,
          ylabel = {[\unit{\giga\flops\double\per\square\milli\meter}]},
          ymajorgrids = true,
          yminorgrids = true,
          minor y tick num = 0,
          ytick style = {draw=none},
          ymin = 0,
          ymax = 30,
          extra y ticks={5, 15, 25},
          xlabel = {Benchmark},
          xmajorgrids = true,
          xtick pos = left,
          xtick = \empty,
          extra x ticks={0, 1, 2, 3},
          extra x tick labels={matmul, conv2d, dotp, fft},
          enlarge x limits = 0.15]

          \pgfplotstableread{results/spatz/ssr}\loadedtable
          \addplot[fill=color2, postaction={pattern=north west lines}] table [x=Index, y expr=\thisrowno{3}/0.73] {\loadedtable};

          \pgfplotstableread{results/spatz/spatz}\loadedtable
          \addplot[fill=color3, postaction={pattern=crosshatch dots}] table [x=Index, y expr=\thisrowno{3}/0.58] {\loadedtable};
        \end{axis}
      \end{tikzpicture}}
    \subcaption{Area Efficiency.}
    \label{fig:area_eff}
  \end{minipage}\hfill%
  \begin{minipage}[h]{0.3\linewidth}
    \resizebox{\linewidth}{!}{\begin{tikzpicture}[/tikz/font=\footnotesize]
        \begin{axis}[
          ybar,
          height = 5cm,
          enlargelimits = 0,
          scaled y ticks = false,
          ylabel = {[\unit{\giga\flops\double\per\watt}]},
          ymajorgrids = true,
          yminorgrids = true,
          minor y tick num = 0,
          ytick style = {draw=none},
          ymin = 0,
          ymax = 100,
          xlabel = {Benchmark},
          xmajorgrids = true,
          xtick pos = left,
          xtick = \empty,
          extra x ticks={0, 1, 2, 3},
          extra x tick labels={matmul, conv2d, dotp, fft},
          enlarge x limits = 0.15]

          \pgfplotstableread{results/spatz/ssr}\loadedtable
          \addplot[fill=color2, postaction={pattern=north west lines}] table [x=Index, y expr=\thisrowno{3}/\thisrowno{5}] {\loadedtable};

          \pgfplotstableread{results/spatz/spatz}\loadedtable
          \addplot[fill=color3, postaction={pattern=crosshatch dots}] table [x=Index, y expr=\thisrowno{3}/\thisrowno{5}] {\loadedtable};
        \end{axis}
      \end{tikzpicture}}
    \subcaption{Energy Efficiency.}
    \label{fig:ener_eff}
  \end{minipage}\hfill%
  \begin{minipage}[h]{0.3\linewidth}
    \resizebox{\linewidth}{!}{\begin{tikzpicture}[/tikz/font=\footnotesize]
        \begin{axis}[
          ybar,
          height = 5cm,
          enlargelimits = 0,
          scaled y ticks = false,
          legend style={
            at={(0.80,0.83)},
            anchor=center,
            legend columns=1
          },
          legend cell align={left},
          ylabel = {[\unit{\giga\flops\double\per\square\milli\meter\per\watt}]},
          ymajorgrids = true,
          yminorgrids = true,
          minor y tick num = 0,
          ytick style = {draw=none},
          ymin = 0,
          ymax = 200,
          xlabel = {Benchmark},
          xmajorgrids = true,
          xtick pos = left,
          xtick = \empty,
          extra x ticks={0, 1, 2, 3},
          extra x tick labels={matmul, conv2d, dotp, fft},
          enlarge x limits = 0.15]

          \pgfplotstableread{results/spatz/ssr}\loadedtable
          \addplot[fill=color2, postaction={pattern=north west lines}] table [x=Index, y expr=\thisrowno{3}/\thisrowno{5}/0.73] {\loadedtable};

          \pgfplotstableread{results/spatz/spatz}\loadedtable
          \addplot[fill=color3, postaction={pattern=crosshatch dots}] table [x=Index, y expr=\thisrowno{3}/\thisrowno{5}/0.58] {\loadedtable};

          \legend{\gls{SSR}, Spatz}
        \end{axis}
      \end{tikzpicture}}
    \subcaption{Combined area/energy Efficiency.}
    \label{fig:area_ener_eff}
  \end{minipage}

    \caption{Efficiency metrics of the Spatz-based and \gls{SSR}-based clusters on a set of data-parallel workloads at nominal conditions (\qty{1}{\giga\hertz}, TT, \qty{0.80}{\volt}, \qty{25}{\celsius}). The considered cluster area does not include the area of the L1 \gls{SPM} \gls{SRAM} macros.}
  \label{fig:efficiencies}
\end{figure*}

We can use other Snitch-based cluster implementations in the literature as proxies to evaluate the area of the Spatz-based cluster. First, the Snitch-based cluster of~\cite{Zaruba2020} was implemented with GlobalFoundries' 22FDX \gls{FDSOI} node, containing \qty{128}{\kibi\byte} of L1 \gls{SPM} divided into \num{32} \gls{SRAM} banks and eight double-precision \glspl{FPU}. The cluster was implemented as a block of dimensions \qtyproduct{858x1046}{\micro\meter}, \qty{0.90}{\milli\meter\squared} in total. However, the cluster's \glspl{FPU} did not have packed-\gls{SIMD} support when running low-precision operations, \ie each \gls{FPU} can only execute one fp64 or fp32 operation per cycle. Therefore, an area comparison between our design and the Snitch cluster~\cite{Zaruba2020} would be biased towards the smaller footprint of the latter.

For a fair comparison, we ported the Snitch cluster~\cite{Zaruba2020} to GlobalFoundries' 12LPP technology with the same frequency target as Spatz. Furthermore, we also added packed-\gls{SIMD} support to the \glspl{FPU} of the Snitch cluster~\cite{Paulin2022,Bertaccini2022}. \Cref{fig:area_eff} shows the area efficiency of the Spatz-based and \gls{SSR}-based clusters on a set of data-parallel workloads. We did not consider the area of the L1 \gls{SPM} \gls{SRAM} macros in the computation. The \gls{SSR}-based cluster~\cite{Paulin2022} is \qtyadj{0.90}{\square\milli\meter} large, with \qty{0.17}{\square\milli\meter} occupied by \qty{128}{\kibi\byte} of L1 \gls{SPM}. The Spatz-based cluster is \qty{0.72}{\square\milli\meter} large, with \qty{0.14}{\square\milli\meter} occupied by also \qty{128}{\kibi\byte} of L1 \gls{SPM}. On the \emph{matmul} kernel, the Spatz-based cluster reaches up to \qty{26.7}{\giga\flops\double\per\square\milli\meter}, \qty{32}{\percent} higher than the efficiency of the \gls{SSR}-based cluster. More strikingly, the Spatz-based cluster reaches an area efficiency of \pchange{\qty{5.9}{\giga\flops\double\per\square\milli\meter}} on \emph{fft} kernel, which is $\num{2.3}\times$ larger than the \pchange{\qty{2.6}{\giga\flops\double\per\square\milli\meter}} area efficiency of the \gls{SSR}-based cluster on the same kernel.

In terms of energy efficiency, \Cref{fig:ener_eff}, the Spatz cluster reaches up to \qty{99.3}{\giga\flops\double\per\watt}. \pchange{Unlike our area efficiency calculations, we did consider the energy consumption of the L1 \gls{SPM} \gls{SRAM} macros since its contribution impacts the optimal $\mathtt{VLENB}$ and the number of memory accesses and the access pattern change when changing the architecture.} In general, Spatz achieves higher energy efficiency than the \gls{SSR}-based cluster for highly compute-intensive workloads such as \emph{matmul} and \emph{conv2d}. For data-intensive workloads such as \emph{dotp}, Spatz reaches \pchange{\qty{48.2}{\giga\flops\double\per\watt}, \qty{30}{\percent}} lower than the energy efficiency of the \gls{SSR}-based cluster on the same workload. The dot product workload, in particular, has very little data reuse. Therefore, transferring data from the L1 \gls{SPM} to the L0 \gls{VRF} does not improve data locality, explaining the energy efficiency drop \pchange{caused by Spatz' current bandwidth to L1 memory, which is lower than the one in the \gls{SSR}-based cluster}. Finally, for the \emph{fft} kernel, Spatz reaches \pchange{\qty{32.7}{\giga\flops\double\per\watt}}, \pchange{\qty{48}{\percent}} larger than the energy efficiency of the \gls{SSR}-based cluster.

Spatz efficiency scales well for lower-precision formats. For example, the Spatz cluster achieves \qty{358}{\giga\flops\half\per\watt} on a \numproduct{128x128} \emph{wid-matmul}$_{16}$ matrix multiplication between fp16 elements. This is \num{3.74}$\times$ higher than the peak efficiency achieved by the \emph{matmul} double-precision kernel, with further energy efficiency gains possible with the ExSdotp~\cite{Bertaccini2022} extension.

Finally, we analyze the cluster's combined area and energy efficiency in \Cref{fig:area_ener_eff}. Thanks to the small footprint and highly-competitive energy efficiency of the Spatz cluster, it reaches up to \qty{171}{\giga\flops\double\per\square\mm\per\watt} on the \emph{conv2d} workload. This is \qty{30}{\percent} larger than the peak area/energy efficiency of the \gls{SSR}-based cluster. This implies that, under the same area constraints, a shared-L1 cluster based on Spatz would reach an energy efficiency \qty{30}{\percent} larger than an \gls{SSR}-based cluster. The combined area/energy efficiency of the Spatz cluster is \pchange{only \qty{12}{\percent}} lower than that of the \gls{SSR}-based cluster for the \emph{dotp} workload. However, for an \emph{fft}, Spatz' \pchange{\qty{56}{\giga\flops\double\per\square\mm\per\watt}} is \pchange{almost twice} the achieved combined area/energy efficiency of the \gls{SSR}-based cluster on the same workload.


\section{Related Work}
\label{sec:related-work}

Increasing memory capacity to lower the bandwidth requirements at a higher level of the memory hierarchy is a fundamental trade-off found in several architectures. For instance, \glspl{GPU} have a large multi-context register file, allowing fast context switching between many concurrent threads. Each \gls{SM} of Nvidia's Hopper \glspl{GPU} has a \qtyadj{256}{\kibi\byte}-large register file to hold the context of up to \num{2048} threads~\cite{NvidiaH1002020}. Therefore, a considerable design effort is dedicated to concurrently optimizing the capacity, bandwidth, reliability, and power consumption of such large register files~\cite{Sparsh2017}. \Cref{tab:relwork} compares our \gls{PPA} metrics against the Snitch cluster~\cite{Zaruba2020}, \change{Vitruvius+~\cite{Minervini2022}}, and Ara~\cite{Perotti2024}, \change{on the largest problem that fits each design's \gls{SPM}}. Spatz is highly competitive in sustained fraction of peak performance, as well as in energy efficiency.

\begin{table}[htbp]
  \centering
  \caption{Comparison between Spatz (us), Snitch~\cite{Zaruba2020}, Vitruvius+~\cite{Minervini2022}, and Ara~\cite{Perotti2024} on an $n \times n$ matrix multiplication.}
  \label{tab:relwork}
  \begin{tabular}[h]{llrrrr}
    \toprule
                    & Unit                                 & \emph{Spatz} & Snitch      & \change{Vitr.}       & Ara         \\\midrule
    Problem size    & $n$                                  & 64           & 64          & \change{256}         & \change{256}         \\
    Node            & [\unit{\nano\meter}]                 & 12           & 12          & \change{22}          & 22          \\
    $V_{\text{DD}}$ & [\unit{\volt}]                       & \num{0.80}   & \num{0.80}  & \change{\num{0.80}}  & \num{0.80}  \\
    Clock (typ)     & [\unit{\giga\hertz}]                 & \num{1.26}   & \num{1.30}  & \change{\num{1.40}}  & \pchange{\num{1.35}}  \\
    Peak DP         & [\unit{\giga\flops\double}]          & \num{20.16}  & \num{20.80} & \change{\num{22.40}} & \pchange{\num{21.6}} \\
    Peak SP         & [\unit{\giga\flops\single}]          & \num{40.32}  & \num{41.60} & \change{---}         & \pchange{\num{43.2}} \\
    Sustained DP    & [\unit{\giga\flops\double}]          & \num{19.74}  & \num{18.26} & \change{\num{21.70}} & \pchange{\num{20.95}} \\
    Sustained SP    & [\unit{\giga\flops\single}]          & \num{38.81}  & \num{33.75} & \change{---}         & \pchange{---} \\
    Total Power DP  & [\unit{\watt}]                       & \num{0.207}  & \num{0.227} & \change{\num{0.459}} & \pchange{\num{0.587}} \\
    Efficiency DP   & [\unit{\giga\flops\double\per\watt}] & \num{97.39}  & \num{92.03} & \change{\num{47.30}} & \pchange{\num{35.70}} \\\bottomrule
  \end{tabular}
\end{table}

At a scale much smaller than \glspl{GPU}, \gls{IOT} end nodes require high performance and energy efficiency under a tight power budget to execute edge-\gls{AI} workloads. In this scenario, the \gls{SIMT} computation paradigm of \glspl{GPU} is hard to leverage, as it requires a massive amount of multi-threaded parallelism (and, therefore, a very large register file). 
Typical \gls{IOT} end nodes exploit the \gls{SIMD} computing paradigm over a moderate amount of functional units. 
For example, Dustin~\cite{Garofalo2021} proposes a configurable \gls{VLEM}, in which its \num{16} \qtyadj{32}{\bit} cores are synchronized, and a single instruction fetch controls all \num{16} datapaths. Therefore, Dustin can achieve the high energy efficiency of packed-\gls{SIMD} architectures while keeping the flexibility of the \gls{MIMD} paradigm. In \gls{VLEM} mode, the register files of the scalar cores act as a single, \qtyadj{512}{\bit}-wide register file. Unlike Spatz, Dustin was designed for low-precision (\eg \qtyadj{2}{\bit}) integer computation. However, its architecture highlights the trend of compact processing elements exploiting \gls{SIMD} computation. 
Intel's \gls{AVX} family of \glspl{ISA} also exploits the packed-\gls{SIMD} paradigm. Its widest \gls{ISA} variant proposes \num{32} \qtyadj{512}{\bit}-wide registers~\cite{Reinders2017} on Intel's line of high-performance processors. With the recent introduction of the AVX10 \gls{ISA}, Intel introduces a common vector \gls{ISA} on the high-performance (P) product lines (\eg Intel Xeon) and the to-be-introduced high-efficiency (E) product line, with a shorter \qtyadj{256}{\bit} vector length on the latter~\cite{IntelAVX2023}.

It is important to note that all packed-\gls{SIMD} solutions tie the register file width to the number of functional units in the system. Thus, the L0 capacity of the \gls{GPRF} is not available as an architectural tuning knob once the \gls{ISA} is defined. The vector-\gls{SIMD} paradigm provides a clean abstraction to separate the \gls{VRF} capacity from the architecture's programming model. Specifically, the same software can run on any vector machine thanks to \gls{VLA} parametrization: programs automatically adjust their execution to exploit the longest possible vector, which is impossible with packed-\gls{SIMD} architectures~\cite{Kozyrakis2003a} and prohibitively expensive with \glspl{GPU}. Therefore, vector processors present the unique characteristic of allowing a full exploration of the memory capacity and bandwidth trade-off. Many vector processing units have been proposed in recent years, thanks to new vector \glspl{ISA} such as Arm's \gls{SVE}~\cite{Armv81M2019} and RISC-V's \gls{RVV}~\cite{RISCV2022}. Examples of such large-scale vector architectures based on the \gls{RVV} \gls{ISA} include BSC's Vitruvius~\cite{Minervini2022}, PULP Platform's Ara~\cite{Perotti2024}, and SiFive's P270~\cite{SiFiveP270}\pchange{, P870~\cite{SiFiveP870},} and X280~\cite{SiFiveX280} cores, to name a few. 

\change{Ultimately, vector processing have historically been associated with supercomputers ~\cite{Russell1978,Yoshida2018}, which limited its application to the \glspl{PE} of a compact shared-L1 cluster. In fact, vector processors usually include all microarchitectural tricks to increase \gls{ILP} and target very long vectors to amortize the \gls{VNB} \cite{10.1145/3624062.3624231} and the massive overheads of complex instruction issue mechanisms. As a result, the energy efficiency of these high-performance architectures is encumbered by their} \pchange{complex, out-of-order} \change{microarchitecture~\cite{Green5002022}.}

However, the defining characteristic of a vector processor is not \gls{ILP} but \gls{DLP}. This key observation led to the design of streamlined vector cores where most hardware resources are dedicated to \gls{DLP} support~\cite{Perotti2024}. In this vein, the idea of an embedded vector machine is gaining traction with modern vector \glspl{ISA}. Arm's \gls{MVE}~\cite{Armv81M2019} and the $\mathtt{Zve*}$ subset of RISC-V's \gls{RVV} \gls{ISA}~\cite{RISCV2022} target small vector machines for edge processing.

Arm's Helium \gls{MVE} is an optional extension proposed as part of the Armv8.1-M architecture~\cite{Armv81M2019}. The Arm Cortex-M55~\cite{ArmCortexM552020} is the first processor to ship with support to \gls{MVE}. However, no quantitative assessment of the performance and efficiency of a Cortex-M55 silicon implementation has been reported in the open literature; hence, a quantitative comparison with Spatz is impossible. For what concerns a qualitative comparison, we observe that the Helium \gls{MVE} defines eight \qtyadj{128}{\bit} wide vector registers as aliases to the floating-point register file. On the M55 processor, the \qtyadj{64}{\bit} datapath means Helium operates on a ``dual-beat regime,'' \ie vector instructions execute in two cycles. This is enough to overlap the execution of successive vector instructions in different processing units without a superscalar core. However, the scalar core must frequently issue instructions to the Helium-capable processing unit to keep its pipeline busy. In contrast, Spatz' longer vector registers and \gls{RVV}'s \gls{LMUL} register grouping allow for a maximum vector length of \num{4096} bits, keeping Spatz busy for \num{32} cycles. This long execution, as shown by our results, massively amortizes the energy overhead of the scalar core, which is a considerable part of the overall energy consumption even on extremely data-parallel kernels such as the matrix multiplication.

In general, architectures based on a tightly-coupled cluster of small vector processors have not been explored in past literature, possibly due to the novelty of small vector machines. Small-scale vector units have been proposed for \glspl{FPGA}, where the leanness of the vector processor is a constraint due to limited \gls{FPGA} resources. Vicuna~\cite{Platzer2021} is a timing-predictable \gls{RVV}-compliant vector processor, synthesized on a Xilinx Series 7 \gls{FPGA}. Its \gls{VRF} was implemented as a multi-ported \gls{RAM} due to concerns with timing anomalies with a multi-banked \gls{VRF}. Vicuna's largest configuration achieves up to \qty{117}{\op\per\cycle} on an \qtyadj{8}{\bit} \numproduct{1024x1024} matrix multiplication kernel. Vicuna's multi-ported \gls{VRF} is similar to Spatz' \gls{VRF}. There is no study about Vicuna's scaling nor an \gls{ASIC} implementation of this architecture, making a power or energy efficiency comparison with Spatz difficult. The same can be said about other small-scale \gls{RVV} vector units demonstrated on \pchange{\glspl{FPGA}~\cite{AlAssir2021, Johns2020, 10.1145/3597031.3597047}}.


\section{Conclusions}
\label{sec:conclusions}

\add{We introduced Spatz, an open-source highly efficient compact \gls{VPU} based on the \gls{RVV} specification.  We couple Spatz with the lightweight Snitch core and optimize its latch-based \gls{VRF} targeting maximum energy efficiency, leveraging the \gls{RVV} \gls{ISA} to boost performance. Spatz differs from typical vector processors, whose \glspl{VRF} have to be much larger to amortize the overhead of typically-used \gls{ILP} techniques (\eg renaming and out-of-order execution).}

\add{We use Spatz as the \gls{PE} of a tightly-coupled shared-L1 cluster, comprising a set of \glspl{PE} sharing tightly-coupled L1 memory through a low-latency interconnect. 
The Spatz cluster contains eight multi-precision \glspl{FPU} and \qty{128}{\kibi\byte} of L1 \gls{SPM}. We implemented this cluster as a \qtyadj{0.74}{\square\milli\meter}-large macro with GlobalFoundries' \gls{FINFET} \qty{12}{\nano\meter} node at an operating frequency of \qty{950}{\mega\hertz} in worst-case conditions (SS, \qty{0.72}{\volt}, \qty{125}{\celsius}). To the best of our knowledge, the dual-core Spatz cluster is the first open-source multi-core \gls{RVV}-based processor architecture.}

\add{We used Kung's architectural balance concept to amortize the energy cost of accessing the L1 \gls{SPM} by tuning the capacity of the highly efficient \gls{VRF} (L0 memory) private to each Spatz, quantitatively showing how vector processors can not only alleviate the \gls{VNB} on the instruction side of the memory but also the memory bottleneck on the data side, thanks to improved data reuse in their \glspl{VRF}. We developed a mathematical model to analyze the energy consumption of the cluster when faced with target workloads and demonstrated its use with a double-precision floating-point matrix multiplication kernel as a function of the \gls{VRF} (L0) size. A very small \gls{VRF} (L0) is needed to balance the L0 and L1 data access costs.}

The Spatz-based cluster achieves high area and energy efficiency on compute-intensive workloads. Spatz achieves an \qty{7.7}{\fma\per\cycle} when running a \numproduct{64x64} double-precision floating-point matrix multiplication, corresponding to \qty{15.7}{\giga\flops\double} and \qty{95.7}{\giga\flops\double\per\watt} at \qty{1}{\giga\hertz} and nominal operating conditions (TT, \qty{0.80}{\volt}, \qty{25}{\celsius}), with more than \qty{55}{\percent} of the power spent on the \glspl{FPU}. Furthermore, the optimally-balanced Spatz-based cluster reaches a \qty{95.0}{\percent} \gls{FPU} utilization (\qty{7.6}{\fma\per\cycle}), \qty{15.2}{\giga\flops\double}, and \qty{99.3}{\giga\flops\double\per\watt} (\qty{61}{\percent} of the power spent in the \gls{FPU}) on a 2D workload with a \numproduct{7x7} kernel, resulting in an outstanding area/energy efficiency of \qty{171}{\giga\flops\double\per\watt\per\square\milli\meter}. At the same area, a computing cluster built upon compact vector processors reaches an energy efficiency \qty{30}{\percent} higher than that of a cluster with the same \gls{FPU} count built upon scalar RISC-V-based cores specialized for stream-based floating-point computation.

This paper shows that vector processors are a sound approach for building \gls{PE} for shared-L1 clusters. \add{The addition of a \gls{VRF} acting as L0 memory allows for a reduction in the bandwidth requirement in the L1 interconnect when the \gls{VRF} (L0) allows for better data reuse, like in key kernels in \gls{AI} such as matrix multiplication.} Given the slowdown of wiring and \gls{SRAM} technology scaling, this reduction simplifies the cluster's low-latency interconnect, the typical factor limiting the physical implementation in current and future technologies.


\section*{Acknowledgment}
\label{sec:acknowledgment}

\pchange{This work was supported in part through the TRISTAN (\#101095947) and the ISOLDE (\#101112274) projects, both funded through the Chips Joint Undertaking (CHIPS JU) of the  European Union’s Horizon Europe’s research and innovation programme and its members.} 


\vskip -1.5\baselineskip plus -1fil
\begin{IEEEbiography}[{\includegraphics[width=1in, height=1.25in, clip, keepaspectratio]{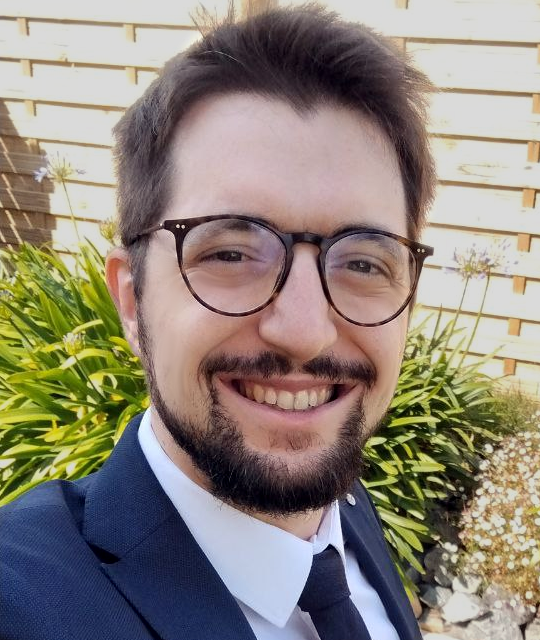}}] {Matteo Perotti} received his M.Sc.\ degree in Electronic Engineering from the Polytechnic University of Turin, Italy, in 2019. He is currently pursuing a Ph.D.\ degree at the Integrated Systems Laboratory of ETH Zurich, Switzerland. His research interests include highly efficient computing architectures and computation with high dynamic-range data types.
\end{IEEEbiography}
\vskip -2.7\baselineskip plus -2fil

\begin{IEEEbiography}[{\includegraphics[width=1in, height=1.25in, clip, keepaspectratio]{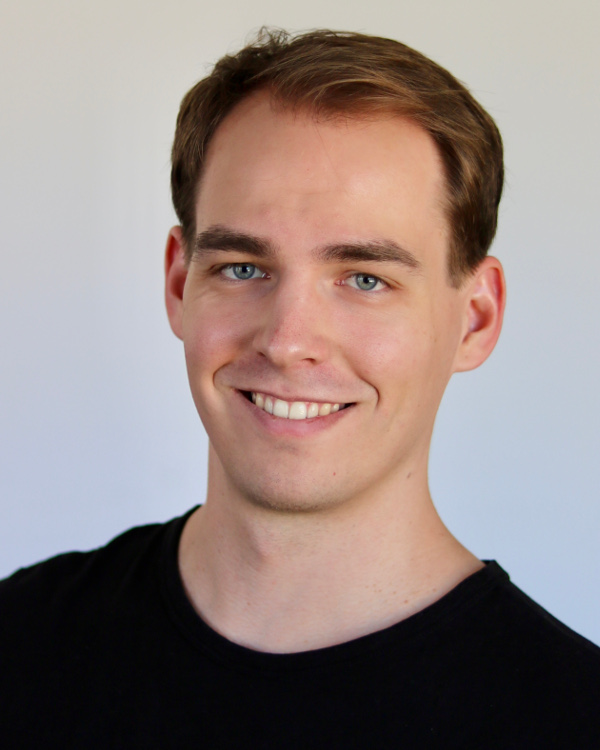}}] {Samuel Riedel} received the B.Sc.\ and M.Sc.\ degree in Electrical Engineering and Information Technology at ETH Zurich in 2017 and 2019, respectively. He is currently pursuing a Ph.D.\ degree with the Digital Circuits and Systems group of ETH Zurich, under the supervision of Prof.\ Luca Benini. His research interests include computer architecture, focusing on many-core systems and their programming model.
\end{IEEEbiography}
\vskip -2.7\baselineskip plus -2fil

\begin{IEEEbiography}[{\includegraphics[width=1in, height=1.25in, clip, keepaspectratio]{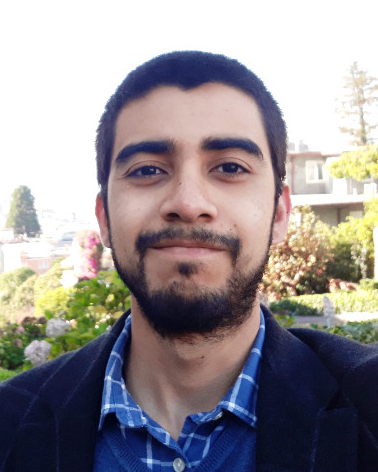}}] {Matheus Cavalcante} received his M.Sc.\ degree in Integrated Electronic Systems from the Grenoble Institute of Technology (Phelma) in 2018, and his Ph.D.\ from ETH Zurich in 2023. During his Ph.D.\ studies, Matheus worked with the Digital Circuits and Systems Group under the supervision of Prof.\ Luca Benini. Matheus' research interests include vector processing, large-scale high-performance computer architectures, and emerging VLSI technologies.
\end{IEEEbiography}
\vskip -2.7\baselineskip plus -2fil

\begin{IEEEbiography}[{\includegraphics[width=1in, height=1.25in, clip, keepaspectratio]{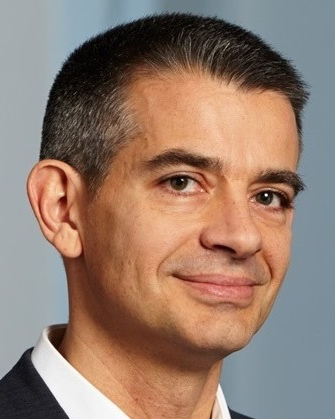}}] {Luca Benini} holds the chair of Digital Circuits and Systems at ETH Zurich and is a Full Professor at the Università di Bologna. He received a Ph.D.\ from Stanford University. He is a Fellow of the ACM and a member of the Academia Europaea. He is the recipient of the 2023 IEEE CS E.\ J.\ McCluskey Award.
\end{IEEEbiography}


\bibliographystyle{IEEEtran}
\bibliography{spatz,ieeetran}

\end{document}